\documentclass[journal,10pt]{IEEEtran}
\ifCLASSINFOpdf
\else
   \usepackage[dvips]{graphicx}
\fi
\usepackage{url}

\usepackage{graphicx}
\usepackage{amsmath}
\usepackage{amssymb}
\usepackage{amsthm}
\usepackage{bbm}
\usepackage{subfigure}
\usepackage{bm}
\usepackage{bbold}
\usepackage[usenames,dvipsnames]{xcolor}
\usepackage{booktabs}
\usepackage{cite}
\allowdisplaybreaks[4]
\usepackage{multirow}
\usepackage{algorithm}
\usepackage{algorithmic}
\usepackage{stmaryrd}

\usepackage{xr}
\makeatletter
\newcommand*{\addFileDependency}[1]{% argument=file name and extension
  \typeout{(#1)}
  \@addtofilelist{#1}
  \IfFileExists{#1}{}{\typeout{No file #1.}}
}
\makeatother

\begin{document}

\title{Low-Complexity RSS-based Underwater Localization with Unknown Transmit Power}

\author{Yingquan Li,~\IEEEmembership{Graduate Student Member,~IEEE}, Jiajie Xu,~\IEEEmembership{Member,~IEEE}, Bodhibrata Mukhopadhyay,~\IEEEmembership{Member,~IEEE}, and Mohamed-Slim Alouini,~\IEEEmembership{Fellow,~IEEE}
\thanks{Y. Li, J. Xu, and M. S. Alouini are with the Computer, Electrical, and Mathematical Sciences and Engineering Division (CEMSE), King Abdullah University of Science and Technology (KAUST), Thuwal, 23955-6900, Kingdom of Saudi Arabia. B. Mukhopadhyay is with the Department of Electronics and Communication, Indian Institute of Technology Roorkee, Uttarakhand, 247667, India.} }

\maketitle

\begin{abstract}
Underwater wireless sensor networks (UWSNs) have received significant attention due to their various applications, with underwater target localization playing a vital role in enhancing network performance. Given the challenges and high costs associated with UWSN deployments, Received Signal Strength (RSS)-based localization offers a viable solution due to its minimal hardware requirements and cost-effectiveness. In this paper, we assign distance-based weights to RSS measurements, providing higher reliability to closer anchor nodes. Using the weighted RSS measurements and generalized trust region subproblem (GTRS), we propose the GTRS-based localization technique with Unknown Transmit Power (GUTP), which can be solved by a simple bisection method. Unlike conventional localization methods that require prior knowledge of the target node's transmit power, GUTP jointly estimates both the location and transmit power of the target node, broadening its practical use. Additionally, we derive the Cramer-Rao lower bounds (CRLBs) for RSS-based underwater localization with known and unknown transmit power, respectively. Extensive simulations demonstrate that GUTP achieves enhanced accuracy and significantly lower computational complexity in estimating the target node's location and transmit power compared to existing semidefinite programming (SDP)-based techniques.
\end{abstract}


\IEEEpeerreviewmaketitle

\section{Introduction}
\subsection{Background and Motivations}
\IEEEPARstart{U}{nderwater} wireless sensor networks (UWSNs) are increasingly significant due to their diverse applications in marine geological research, environmental monitoring, and underwater archaeology~\cite{Review, Xue2, Inam}. In underwater environments, signal propagation is subject to significantly greater attenuation than in terrestrial settings, due to the additional absorption introduced by the water medium~\cite{Intro_underwater1, Intro_underwater2, Xue1}. Consequently, traditional high-frequency radio communication techniques encounter severe limitations in underwater scenarios. For instance, the commonly employed 2.4~GHz Wi-Fi signals exhibit attenuation levels as high as 1000~dB/m in seawater~\cite{Intro_RF1}. Even for low-frequency carriers at 10~MHz, attenuation remains substantial, reaching approximately 100~dB/m in saline water~\cite{Intro_RF2}. Although specialized underwater sensors have been developed to enhance the detection of weak RF signals, their performance degrades rapidly with increasing transmission distance~\cite{Intro_RF3}. Accordingly, practical underwater RF communication is constrained to short-range, near-surface, or shallow-water scenarios, typically with effective ranges of less than one meter~\cite{Intro_RF4, WCM}. In recent years, underwater optical wireless communication (UOWC) has garnered increasing attention, owing to the existence of a low-attenuation window in the blue-green spectrum where absorption is minimized~\cite{Intro_UOWC1, Intro_UOWC2, Xue3}. Experimental studies have demonstrated the potential of UOWC in moderate-range applications. For instance, the authors in~\cite{Intro_UOWC2} achieved a stable data rate of 125~Mb/s over a 30~m link maintained continuously for 30 days, while \cite{Intro_UOWC3} reported a packet loss rate of 32\% over a 7.5~m link at 0.2~Mb/s in water with a turbidity of 25~nephelometric turbidity units (NTU). Despite these advances, the applicability of optical signals remains limited to short distances, which is usually within 100~m. This constraint results from the high power requirements for long-distance transmission, which exceed the current capabilities in the fabrication of optical sources~\cite{Intro_UOWC4, Intro_UOWC5}. However, in UWSNs, nodes are commonly deployed in deep-water environments, typically at depths greater than 1~km~\cite{Intro_UWSN1}. In addition, the high cost of deployment leads to sparse node distribution, with inter-node distances spanning several kilometers~\cite{Intro_underwater1, Intro_UWSN2}. In this context, acoustic signals are widely used to connect nodes within UWSNs, as they exhibit low attenuation, support high transmit power, and enable non-line-of-sight (NLOS) propagation~\cite{SDPXu, SDSOCP, urick1975principles}.

A UWSN consists of multiple anchor nodes with known locations and one target node with an unknown location. A key aspect of enhancing the effectiveness of these networks is the accurate localization of the underwater target node, as node's location directly influence the analysis of collected data~\cite{FCUP}. Furthermore, localizing unknown underwater target node can expand the application scenarios of UWSNs in both civilian and military sectors. However, compared to terrestrial localization, underwater localization presents unique challenges due to complex network deployment, sophisticated signal propagation models, and high maintenance costs~\cite{U1, U3, SDSOCP}. These factors necessitate low-complexity and cost-effective localization techniques, making received signal strength (RSS)-based localization a promising solution due to its straightforward hardware architecture and ease of implementation.

\subsection{Related Works}\label{subsec:Intro_RelatedWork}
RSS-based localization estimates the distance between the target node and anchor nodes using RSS measurements from target-anchor links, enabling the determination of the target node's location through coordination with multiple anchor nodes. Most RSS-based underwater localization techniques assign equal priority to all target-anchor links. Xian~\textit{et al.}~\cite{CM_S1} introduces a robust coarse-to-fine localization algorithm (RCFLA) that accounts for uncertainty in the target node's transmit power and a time-varying path loss exponent (PLE). The algorithm initiates with a rough estimation using the active set method (ASM), followed by an iterative Broyden-Fletcher-Goldfarb-Shanno (BFGS) trust-region method to enhance convergence toward the global optimum. They also provided a theoretical analysis of the convergence behavior of the proposed technique. However, the error in distance estimation based on RSS measurements is dependent on the actual inter-node distance; specifically, for a given deviation in RSS measurements, longer-range links tend to exhibit larger distance estimation errors compared to shorter-range links~\cite{mag, Weighted1}. Consequently, higher reliability should be assigned to RSS measurements from shorter links to enhance localization accuracy, as these provide more precise information about the target node's location~\cite{Weighted2}.

Due to the dynamic and complex nature of the underwater environment, achieving precise configuration is challenging. To address this issue, researchers have proposed underwater localization techniques based on artificial intelligence (AI), which reduce reliance on extensive system calibration, thereby enabling more adaptive and robust solutions suitable for real-time deployment~\cite{Intro_AI1, Intro_AI2, R_AI}. In~\cite{Intro_AI1}, the authors proposed an underwater localization technique based on deep learning. Estimated distances from the target node to nearby anchor nodes were obtained using cross-correlation and then provided as input to a convolutional layer. The used neural network outputted the estimated range and direction-of-arrival (DOA) of the target node. Numerical simulation were performed to examine the robustness of their proposed technique in the presence of inaccurately estimated sample delays. The authors in~\cite{R4C1_1} proposed a layer-by-layer watchman-based collision-free routing scheme to improve reliability and mitigate the void hole problem in UWSNs. The authors in~\cite{Intro_AI2} developed a recurrent neural network (RNN) to predict the relative horizontal velocities of an AUV. They trained the RNN using experimental data from an inertial measurement unit, pressure sensor, and control inputs, where a doppler velocity logger provided ground-truth velocities. The predictions of the relative velocities were implemented in a dead-reckoning algorithm to approximate north and east positions. In~\cite{R4C1_2}, the authors proposed an energy-efficient watchman-based flooding algorithm to mitigate void holes and enhance network performance in UWSNs. The algorithm is verified using the Z-Eves toolbox, showing improvements in packet delivery and throughput compared with benchmark protocols. Ali~\textit{et al.}~\cite{R4C1_3} proposed a subnet based hotspot algorithm (SBHA) to partition large-scale dynamic networks into subnets, thereby reducing hotspot occurrence and lowering traffic load near sink nodes. The proposed algorithm is validated through formal specifications using the VDM-SL toolbox to demonstrate improvements in end-to-end delay and network lifetime. The authors in~\cite{R4C1_4} proposed an Energy-Efficient Mobility-Based Watchman Algorithm (E2-MBWA) that employs a layer-by-layer forwarding mechanism to mitigate the hotspot problem in UWSNs, thereby enhancing packet delivery ratio and addressing data storage challenges. The algorithm integrates the Data Packet Forwarding Algorithm (DFPA) with the Watchman Layer Update Mechanism (WLUM) and leverages secondary nodes as substitutes. In~\cite{R4C1_5}, Ali~\textit{et al.} developed a priority ranking algorithm for localization (PRAL), which used an energy-efficient beacon node ranking framework that integrates a modified page rank mechanism with bio-inspired metaheuristic techniques to enhance localization accuracy and optimize data routing in UWSNs. The methodology incorporates ant colony optimization for path selection, artificial bee colony for high-fitness node identification, and fish school search for optimal node selection. In~\cite{R_AI}, the authors presents a RSS-based localization technique for underwater visible light communication (UVLC) systems that integrates Kalman filtering (KF) and optimized deep learning models. They evaluate the combination exponential arbitrary power function (CEAPF) as the channel model and employ an automatic hyperparameter optimization approach based on Bayesian methods to enhance localization performance. Draz~\textit{et al.}~\cite{R4C1_6} proposed a hybrid underwater localization communication framework that integrates blockchain technology with a geometric distance-based localization routing scheme to enhance security and improve data delivery in UWSNs. The methodology evaluates the proposed framework under varying node configurations and mobility scenarios through comparative experiments against state-of-the-art techniques, demonstrating improvements in end-to-end delay and routing efficiency.

\begin{figure*}[!t]
    \centering
    \includegraphics[width=.9\linewidth]{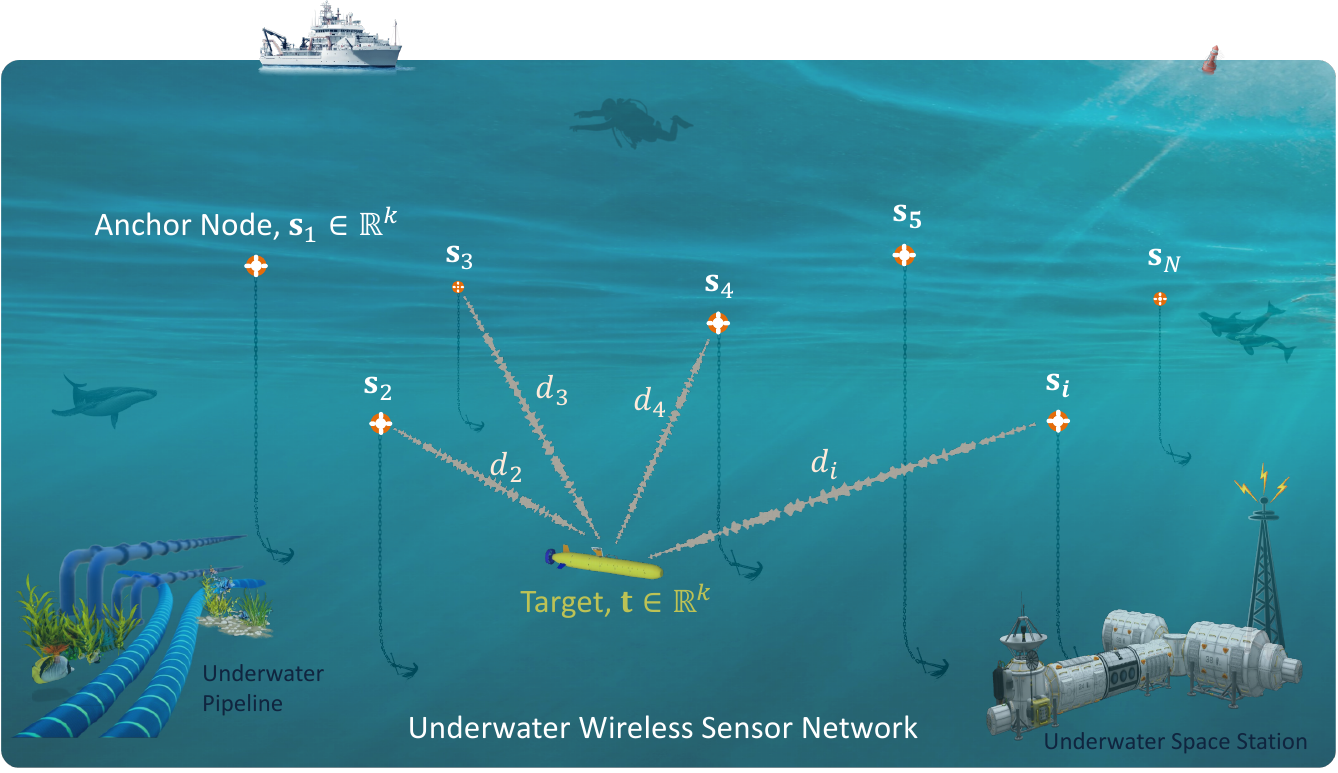}
    \caption{Network architecture of the UWSN for target localization.}
    \label{Fig:Fig1}
\end{figure*}
\subsection{Contributions}\label{subsec:Intro_Contributions}
This paper presents an RSS-based underwater localization technique designed for scenarios where the target node's transmit power is unknown. We use Taylor expansion and least squares (LS) with weighted RSS measurements to propose the generalized trust region subproblem (GTRS)-based localization technique with unknown transmit power (GUTP).
The main contributions are as follows: 
\begin{itemize}
    \item We propose GUTP, a localization technique based on the generalized trust region subproblem (GTRS) framework. GUTP enables efficient solution via a straightforward bisection method. Compared to existing techniques, GUTP exhibits high localization accuracy with significantly lower computational complexity. To the best of our knowledge, this is the first study to apply the GTRS framework with weighted RSS measurements to underwater localization in scenarios with unknown transmit power.
    \item To enhance localization accuracy, we incorporate weighted RSS measurements across different target-anchor links, where higher weights are assigned to anchor nodes in closer distance to the target. The location and transmit power of the target node are estimated jointly.
    \item We derive the CRLB for RSS-based underwater localization in scenarios where the transmit power is both known and unknown.
    \item Extensive simulation results demonstrate that GUTP achieves improved localization performance. By leveraging a simple bisection method to solve the resulting GTRS problem, GUTP significantly reduces computational complexity compared to existing underwater localization techniques.
\end{itemize}

\section{Channel model}\label{Sec:ChannelModel}
To facilitate the development of a low-complexity underwater localization framework, we begin by presenting the system model that characterizes underwater signal propagation. In this section, we outline the formulation of the RSS measurement model in underwater environments, which is the foundation for subsequent derivation. Additionally, we introduce key model parameters that are also significant for deriving the performance bounds discussed in Section~\ref{sec:CRLB_manuscript}. Consider a $k$-dimensional UWSN, as presented in Fig.~\ref{Fig:Fig1}, comprising a target node whose position is estimated based on its communication with $N$ neighboring anchor nodes. Let $\mathbf{t} \in \mathbbm{R}^k$ be the unknown location of the target node and $\mathbf{s}_i \in \mathbbm{R}^k$ be the known location of the $i^{\text{th}}$ anchor node. We assume no failures occur in the network during the localization process, ensuring that all anchor nodes can communicate with the target node. Consequently, the set of indices for the available anchor nodes is $\mathcal{S} = \{1,\dots,N\}$. When the target node functions as the signal source, the transmission loss model for underwater acoustic signals indicates that the RSS measurement at $\mathbf{s}_i$ (in dBm) is given by~\cite{CM_S1,CM_S2}:
\begin{IEEEeqnarray}{lCr}
\label{Eq:CM}
P_i = P_t - 10\beta \log_{10}\frac{d_i}{d_0} - \alpha(d_i-d_0) + n_i,\ i\in\mathcal{S},
\end{IEEEeqnarray}
where $P_i$ is the received power at $\mathbf{s}_i$, while the parameter $\beta$ corresponds to the PLE, $\alpha$ denotes the attenuation factor, and $d_{0}$ is the reference distance. In~\eqref{Eq:CM}, $P_t$ (in~dBm) denotes the RSS measurement at $d_0$ ($d_0\leq d_i$). In this paper, we set $d_0$ to 1~m for simplicity and refer to $P_t$ as the transmit power of $\mathbf{t}$. We denote the Euclidean distance between $\mathbf{t}$ and $\mathbf{s}_i$ as $d_i = \left\|\mathbf{t}-\mathbf{s}_i\right\|$, where $\|\cdot\|$ represents the $\ell_2$--norm. $n_i$ denotes the noise term modeled as a zero-mean Gaussian random variable with variance $\sigma_i^2$. Without loss of generality, we assume that $n_i$ (for $i\in\mathcal{S}$) are mutually independent~\cite{CTUP}. Unlike terrestrial networks, which primarily account for slow fading, underwater channels also experience signal propagation loss due to absorption. This loss is modeled as the product of the propagation distance (in m) and $\alpha$ (in dB/m), given by~\cite{SDPXu,Absorption}:
\begin{IEEEeqnarray}{lCr}
\label{Eq:Absorption}
\alpha = \left(\frac{0.11f^2}{1+f^2}+\frac{44f^2}{4100+f^2}+\frac{2.75f^2}{10000}+0.003\right)\times10^{-3},\IEEEeqnarraynumspace
\end{IEEEeqnarray}
where $f$ (in~kHz) is the center frequency of the transmitted signal. As the communication distance increases from short to long, the frequency $f$ used in UWSNs typically ranges from 1 to 50~kHz. As indicated by~\eqref{Eq:CM}, RSS-based localization requires prior knowledge of $P_t$, $\beta$, and $\alpha$ to estimate the distance between the target node and anchor nodes. However, this information is usually unknown in practice, particularly in scenarios where the targets are hostile, making it challenging to determine the target node's location.

\section{Weighted RSS measurements}\label{Sec:Weighted}
\begin{figure}[!t]
    \centering
    \includegraphics[width=\linewidth]{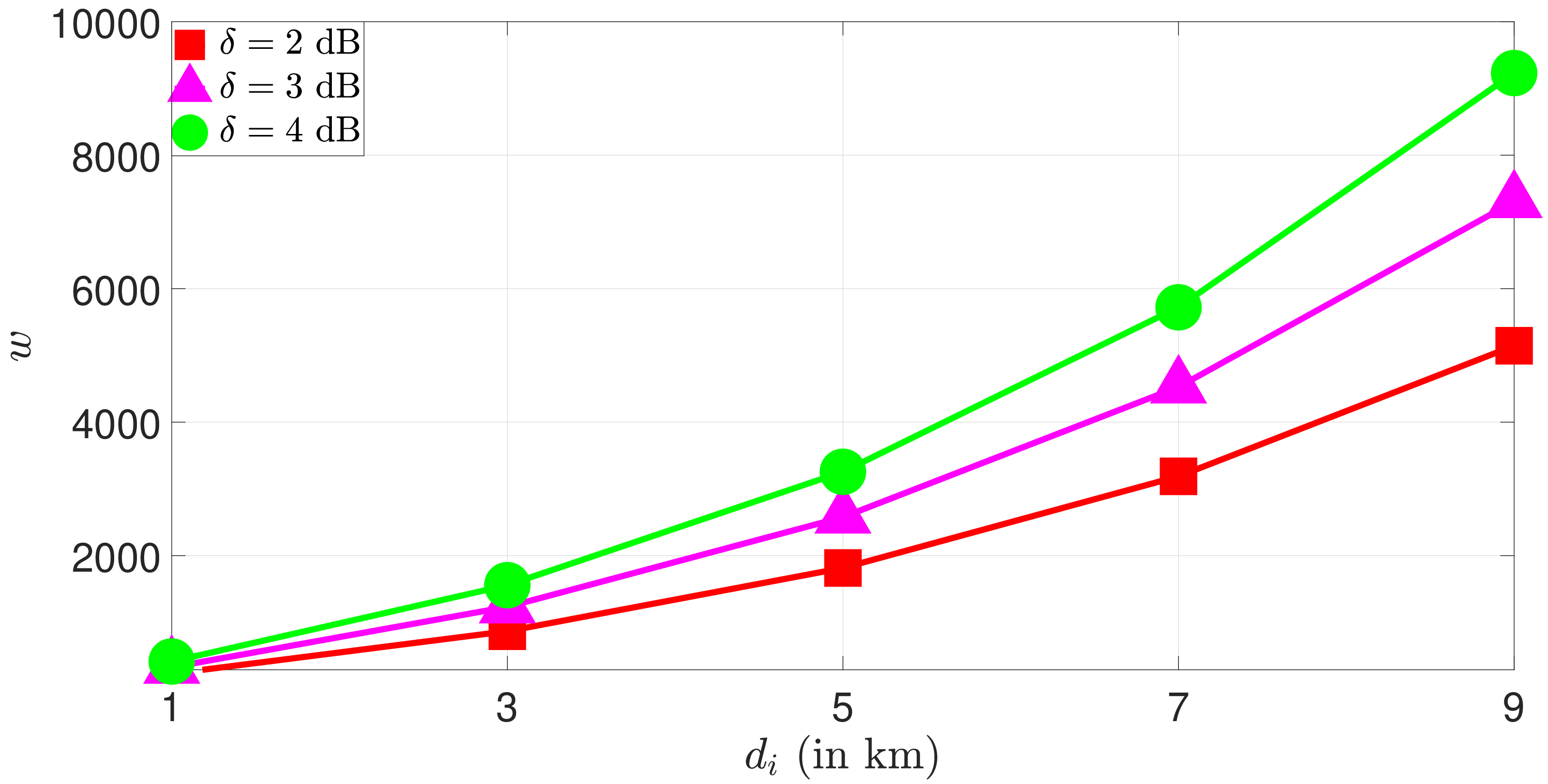}
    \caption{Illustration of the effect of link distance on distance estimation errors under the same noise level.}
    \label{Fig:Weights}
\end{figure}
Since RSS-based ranging does not require additional hardware, we can readily extract the distance information between the target and anchor nodes from the RSS measurement. In this section, we demonstrate that under the same noise level, longer links lead to more significant location estimation errors, thereby justifying the need to assign higher weights to short-range links. By dividing both sides of~\eqref{Eq:CM} by $10\beta$ and then taking the exponent, we obtain:
\begin{IEEEeqnarray}{lCr}
\label{Eq:ydi}
\underbrace{\frac{d_i10^{\frac{\alpha d_i}{10\beta}}}{10^{\frac{P_t}{10\beta}}}10^{\frac{-n_i}{10\beta}}}_{y(d_i,P_t,n_i)} = 10^{\frac{-P_i+\alpha}{10\beta}},
\end{IEEEeqnarray}
where \( y(d_i, P_t, n_i) \) is a function involving estimates of unknown parameters, derived from the known values of \( P_i \), \( \alpha \), and \( \beta \). When \( n_i = 0 \), the scenario is noise-free and yields the best localization performance. The difference between \( y(d_i, P_t, n_i) \) and its value in the noiseless case presents the degradation in performance caused by noise, with larger differences indicating more significant degradation. Assuming $n_i=\delta$ in one sampling, the difference between $y(d_i, P_t, n_i)$ from the erroneous RSS measurement and the noise-free RSS measurement is:
\begin{IEEEeqnarray}{lCr}
\label{Eq:wdi}
\left|y(d_i,P_t,\delta)-y(d_i,P_t,0)\right| = \underbrace{\frac{d_i10^{\frac{\alpha d_i}{10\beta}}}{10^{\frac{P_t}{10\beta}}}\Big(\big|10^{\frac{-\delta}{10\beta}}-1\big|\Big)}_{w(d_i,P_t,\delta)}.
\end{IEEEeqnarray}
We plot the behavior of $w(d_i,P_t,\delta)$ as $d_i$ increases from 1~km to 10~km in Fig.~\ref{Fig:Weights}, with $\beta=2$ representing spherical spreading. The center frequency is set to 9~kHz, resulting in $\alpha = 9.86\times10^{-4}$~dB/m. We assume $P_t = 0$~dBm and set the noise measurement $\delta$ to 2~dB, 3~dB, and 4~dB, respectively. Fig.~\ref{Fig:Weights} presents that for the same link distance, an increase in $\delta$ results in a corresponding rise in $w(d_i,P_t,\delta)$, demonstrating a higher location estimation error. It can be observed that, for the same $\delta$, a larger distance between nodes leads to a more significant difference between $y(d_i,P_t,\delta)$ and $y(d_i,P_t,0)$, thereby increasing location estimation errors. Consequently, we assign a weight to each link as: 
\begin{IEEEeqnarray}{lCr}
\label{Eq:weights}
w_i = \frac{\sum_{i\in\mathcal{S}}10^{\frac{-P_i+\alpha}{10\beta}} - 10^{\frac{-P_i+\alpha}{10\beta}}}{(N-1)\sum_{i\in\mathcal{S}}10^{\frac{-P_i+\alpha}{10\beta}}},\ i\in\mathcal{S}.
\end{IEEEeqnarray}

Unlike~\cite{Weighted1}, we normalize the weights to mitigate numerical instability. By applying weights as presented in~\eqref{Eq:weights} to the RSS measurements, we prioritize those collected from nearby links, subsequently improving the localization accuracy. We then use the weighted RSS measurements to propose a localization technique with unknown transmit power.

\section{Localization technique with unknown transmit power}\label{sec:NTP}
In practice, UWSNs are usually not comprehensively tested or calibrated to reduce implementation costs and simplify deployment. Therefore, it is significant to study cases where the target node's transmit power is not pre-recorded. In this section, we present an RSS-based localization technique addressing scenarios with unknown transmit power. The composition of the logarithmic and norm functions in~\eqref{Eq:CM} is discontinuous at $\mathbf{t} = \mathbf{s}_i$, since $\| \mathbf{t} - \mathbf{s}_i \|$ becomes zero. As a result, $d_i = 0$, causing $\log_{10}d_i$ to diverge toward negative infinity. To mitigate the discontinuity caused by the composition of the logarithmic and norm functions, we take the base-10 exponent on both sides of~\eqref{Eq:CM} and reformulate the terms as:
\begin{IEEEeqnarray}{lCr}
\label{Eq:np1}
d_i10^{\frac{P_i-\alpha}{10\beta}} = 10^{\frac{P_t+n_i}{10\beta}}10^{\frac{-\alpha d_i}{10\beta}}.
\end{IEEEeqnarray}
The absorption coefficient \( \alpha \) given in~\eqref{Eq:Absorption} remains small at low signal frequencies. For instance, at \( f = 9\,\mathrm{kHz} \), \( \alpha = 9.86 \times 10^{-4}\,\mathrm{dB/m} \). Consequently, we can apply a first-order Taylor expansion as follows:
\begin{IEEEeqnarray}{lCr}
\label{Eq:np2}
d_i10^{\frac{P_i-\alpha}{10\beta}} = 10^{\frac{P_t+n_i}{10\beta}}\big(1-\frac{\ln10}{10\beta}\alpha d_i\big).\IEEEeqnarraynumspace
\end{IEEEeqnarray}
Therefore, we can transform~\eqref{Eq:np2} into an equivalent form:
\begin{IEEEeqnarray}{lCr}
\label{Eq:np3}
d_i = \frac{10\beta}{\alpha\ln10} - \frac{10\beta}{\alpha\ln10}\frac{q_i}{q_i+g_i},
\end{IEEEeqnarray}
where $q_i = 10^{\frac{P_i-\alpha}{10\beta}}$ and $g_i = 10^{\frac{P_t+n_i}{10\beta}}\frac{\ln10}{10\beta}\alpha$. For sufficiently small $n_i$ and $\alpha$, we have $\frac{1}{q_i+g_i}\approx\frac{q_i-g_i}{q_i^2}$~\cite{CTUP}, which yields:
\begin{IEEEeqnarray}{cll}
    \label{Eq:np4}
    d_i = \frac{10^{\frac{P_t+n_i}{10\beta}}}{q_i}.
\end{IEEEeqnarray}
By squaring both sides and rearranging terms of~\eqref{Eq:np4}, we obtain:
\begin{IEEEeqnarray}{cll}
    \label{Eq:np5}
    d_i^2q_i^2 = 10^{\frac{P_t}{5\beta}}10^{\frac{n_i}{5\beta}}.
\end{IEEEeqnarray}
For sufficiently small $n_i$, we apply the Taylor expansion to convert~\eqref{Eq:np5} into:
\begin{IEEEeqnarray}{cll}
    \label{Eq:np6}
    d_i^2q_i^2 = 10^{\frac{P_t}{5\beta}}\big(1+\frac{\ln10}{5\beta}n_i\big).
\end{IEEEeqnarray}
We let $\zeta_i = 10^{\frac{P_t}{5\beta}}n_i$ to denote the Gaussian random variable on the right side of~\eqref{Eq:np6}. Therefore, by introducing the auxiliary variable $u = 10^{\frac{P_t}{5\beta}}$ and using $d_i = \|\mathbf{t}-\mathbf{s}_i\|$, we formulate a weighted LS-based estimator given by:
\begin{IEEEeqnarray}{cll}
    \label{Eq:np7}
    \mathop{\rm{min}}_{\mathbf{t},u} & \sum_{i\in\mathcal{S}}w_i\left(\frac{5\beta}{\ln10} q_i^2\|\mathbf{t}\|^2 -\frac{10\beta}{\ln10} q_i^2\mathbf{s}_i^T\mathbf{t} - \frac{5\beta}{\ln10}u \right.\nonumber\\
    &\hspace{4.6cm}\left. + \frac{5\beta}{\ln10} q_i^2\|\mathbf{s}_i\|^2 \right)^2,\IEEEeqnarraynumspace
\end{IEEEeqnarray}
where $w_i$ is presented in~\eqref{Eq:weights}. By solving~\eqref{Eq:np7}, we can jointly estimate the target node's location and transmit power. To facilitate this, we reformulate~\eqref{Eq:np7} in a matrix form as:
\begin{IEEEeqnarray}{cll}
    \label{Eq:estimator_np}
    \IEEEyesnumber\IEEEyessubnumber*
    \mathop{\rm{min}}_{\mathbf{z}} & \  \|\mathbf{W}(\mathbf{R}\mathbf{z}-\mathbf{v})\|^2 \IEEEeqnarraynumspace\\
{\text{s.t.}} 
    & \mathbf{z}^T\mathbf{H}\mathbf{z} + 2\mathbf{h}^T\mathbf{z} = 0,
\end{IEEEeqnarray}
where\footnote{$\mathbf{I}_M$ denotes the identity matrix of order $M$, and $\mathbf{0}_{M\times N}$ represents the $M\times N$ matrix of all zeros.} $\mathbf{z} = [\mathbf{t}^T,\|\mathbf{t}\|^2,u]^T \in \mathbbm{R}^{(k+2)\times 1}$ and $\mathbf{W}$ denotes a diagonal matrix whose $i^{\mathrm{th}}$ diagonal entry is $w_i$, and
\begin{IEEEeqnarray}{lll}
    \label{Eq:elements_np}
    \mathbf{R} = \begin{bmatrix}
                    -\frac{10\beta}{\ln10} q_1^2\mathbf{s}_1^T & \frac{5\beta}{\ln10} q_1^2 & -\frac{5\beta}{\ln10} \\
                    \vdots & \vdots & \vdots & \\
                    -\frac{10\beta}{\ln10} q_i^2\mathbf{s}_i^T & \frac{5\beta}{\ln10} q_i^2 & -\frac{5\beta}{\ln10} \\
                    \vdots & \vdots & \vdots & \\
                    -\frac{10\beta}{\ln10} q_N^2\mathbf{s}_N^T & \frac{5\beta}{\ln10} q_N^2 & -\frac{5\beta}{\ln10}
                 \end{bmatrix},\nonumber\\
    \mathbf{v} = \begin{bmatrix}
                    -\frac{5\beta}{\ln10} q_1^2\|\mathbf{s}_1\|^2, \dots, -\frac{5\beta}{\ln10} q_N^2\|\mathbf{s}_N\|^2 \\
                 \end{bmatrix}^T,\nonumber\\
    \mathbf{H} =  \begin{bmatrix}
                    \mathbf{I}_k & \mathbf{0}_{k\times2} \\
                    \mathbf{0}_{2\times k} & \mathbf{0}_{2\times2}
                 \end{bmatrix},\ \ 
    \mathbf{h} = \begin{bmatrix}
                    \mathbf{0}_{k\times 1} \\
                    -\frac{1}{2} \\
                    0
                    \end{bmatrix}.\IEEEeqnarraynumspace
\end{IEEEeqnarray}

Optimization problems of the form~\eqref{Eq:estimator_np} are referred to as GTRS, which are non-convex~\cite{Bisection}. However, the necessary and sufficient condition for their optimal solution is that there exists a Lagrangian multiplier $\lambda$ such that\footnote{For a symmetric matrix $\mathbf{X}$, $\mathbf{X}\succcurlyeq\mathbf{0}$ implies that $\mathbf{X}$ is positive semidefinite.}~\cite{GTRS2}: 
\begin{IEEEeqnarray}{cll}
    \label{Eq:NScondition}
    \IEEEyesnumber\IEEEyessubnumber*
    (\mathbf{R}^T\mathbf{R} + \lambda\mathbf{H})\mathbf{z} = \mathbf{R}^T\mathbf{v} - \lambda\mathbf{h},\\
    \mathbf{z}^T\mathbf{H}\mathbf{z} + 2\mathbf{h}^T\mathbf{z} = 0,\\
    \mathbf{R}^T\mathbf{R} + \lambda\mathbf{H} \succcurlyeq\mathbf{0}.
\end{IEEEeqnarray}
Consequently, the optimal solution of~\eqref{Eq:estimator_np} is expressed as:
\begin{IEEEeqnarray}{cll}
    \label{Eq:solution_kp}
    \mathbf{z} = (\mathbf{R}^T\mathbf{R} + \lambda\mathbf{H})^{-1}(\mathbf{R}^T\mathbf{v} - \lambda\mathbf{h}),
\end{IEEEeqnarray}
where $\lambda$ is the unique solution to $\mathbf{z}^T\mathbf{H}\mathbf{z} + 2\mathbf{h}^T\mathbf{z}=0$ over the interval $(-\frac{1}{\lambda^{\ast}},\infty)$, with $\mathbf{z}$ defined as in~\eqref{Eq:solution_kp} and $\lambda^{\ast}$ representing the largest eigenvalue of $(\mathbf{R}^T\mathbf{R})^{-\frac{1}{2}}\mathbf{H}(\mathbf{R}^T\mathbf{R})^{-\frac{1}{2}}$. Mor$\Acute{\text{e}}$~\textit{et al.}~\cite{GTRS2} established that $\mathbf{z}^T\mathbf{H}\mathbf{z} + 2\mathbf{h}^T\mathbf{z}$ exhibits strict monotonic decrease within this interval. As a result, we can adopt the bisection method as an efficient approach to determining $\lambda$, which is then used to compute $\mathbf{z}$. We extract the estimates for the target node's location and transmit power using $\hat{\mathbf{t}} = \hat{\mathbf{z}}_{1:k}$ and $\hat{P}_0 = 5\beta\log_{10}\hat{\mathbf{z}}_{k+2}$. We refer to the technique as GUTP.

\section{Cramer-Rao Lower Bound (CRLB)}\label{sec:CRLB_manuscript}
The CRLB provides a theoretical lower bound on the variance of unbiased estimators and is commonly used as a performance benchmark for localization techniques, as it defines the best achievable accuracy under given conditions. However, CRLB in underwater scenarios has not been extensively studied, particularly in cases where $P_t$ is unknown. In this section, we derive the CRLB for RSS-based underwater localization for scenarios with known and unknown $P_t$, respectively. Let $\mathbf{p} = [P_1,\dots,P_N]^T$ and $\bm{\theta} = [\mathbf{t}^T,P_t]^T$ stack RSS measurements and unknown parameters. Therefore, we can derive the log-likelihood of $\mathbf{p}$ given $\bm{\theta}$ from~\eqref{Eq:CM}:
\begin{IEEEeqnarray}{lCr}
    \label{Eq:PDF}
     \ln p\left(\mathbf{p};\bm{\theta}\right) = -\frac{1}{2}\sum_{i\in\mathcal{S}}\ln(2\pi\sigma_i^2)+\nonumber\\
     \sum_{i\in\mathcal{S}}\frac{(P_i-P_t+10\beta\log_{10}\|\mathbf{t}-\mathbf{s}_i\|+\alpha\|\mathbf{t}-\mathbf{s}_i\|-\alpha)^2}{-2\sigma_i^2}.\IEEEeqnarraynumspace
\end{IEEEeqnarray}
The Hessian matrix of $\ln p\left(\mathbf{p};\bm{\theta}\right)$ with respect to $\bm{\theta}$ is given by:
\begin{IEEEeqnarray}{ll}
    \label{Eq:GH}
    \IEEEyesnumber\IEEEyessubnumber*
    \label{Eq:Hessian_t}
    \frac{\partial^2 \ln p\left(\mathbf{p};\bm{\theta}\right)}{\partial \mathbf{t}^2} &= \sum_{i\in\mathcal{S}}\frac{-1}{\sigma_i^2}\left(\frac{\mathbf{c}_i\mathbf{c}_i^T+\ln10\left\|\mathbf{t}-\mathbf{s}_i\right\|^2f_i\mathbf{D}_i}{(\ln10)^2\left\|\mathbf{t}-\mathbf{s}_i\right\|^4} - \right.\nonumber\\
    &\hspace{2.6cm} \left.\frac{2\ln10f_i\mathbf{c}_i(\mathbf{t}-\mathbf{s}_i)^T}{(\ln10)^2\left\|\mathbf{t}-\mathbf{s}_i\right\|^4}\right),\IEEEeqnarraynumspace\\
    \label{Eq:Hessian_t_p}
    \frac{\partial^2 \ln p\left(\mathbf{p};\bm{\theta}\right)}{\partial \mathbf{t}\partial P_t} &= \sum_{i\in\mathcal{S}}\frac{\mathbf{c}_i}{\sigma_i^2\ln10\|\mathbf{t}-\mathbf{s}_i\|^2},\IEEEeqnarraynumspace\\
    \label{Eq:Hessian_p}
    \frac{\partial^2 \ln p\left(\mathbf{p};\bm{\theta}\right)}{\partial P_t^2} &= \sum_{i\in\mathcal{S}}\frac{-1}{\sigma_i^2},\IEEEeqnarraynumspace
\end{IEEEeqnarray}
where
\begin{IEEEeqnarray}{cl}
    \label{Eq:f1f2D}
    \IEEEyesnumber\IEEEyessubnumber*
    \label{Eq:f1}
    f_i &= P_i-P_t+10\beta\log_{10}\|\mathbf{t}-\mathbf{s}_i\|+\alpha\|\mathbf{t}-\mathbf{s}_i\|-\alpha,\IEEEeqnarraynumspace\\
    \label{Eq:f2}
    \mathbf{c}_i &= 10\beta(\mathbf{t}-\mathbf{s}_i) + \alpha\ln10\|\mathbf{t}-\mathbf{s}_i\|(\mathbf{t}-\mathbf{s}_i),\IEEEeqnarraynumspace\\
    \label{Eq:D}
    \mathbf{D}_i &= 10\beta\mathbf{I}_k+\alpha\ln10(\left\|\mathbf{t}-\mathbf{s}_i\right\|\mathbf{I}_k+\frac{(\mathbf{t}-\mathbf{s}_i)(\mathbf{t}-\mathbf{s}_i)^T}{\left\|\mathbf{t}-\mathbf{s}_i\right\|}).\IEEEeqnarraynumspace
\end{IEEEeqnarray}
Since $n_i$ has a zero mean value, we have $\mathbbm{E}[f_i] = 0$, leading to the Fisher information matrix (FIM) for $\bm{\theta}$ as:
\begin{IEEEeqnarray}{llr}
    \label{Eq:Fisher}
    \mathbf{F} &= \begin{bmatrix}
                    -\mathbbm{E}\big[\frac{\partial^2 \ln p\left(\mathbf{p};\bm{\theta}\right)}{\partial \mathbf{t}^2}\big] & -\mathbbm{E}\big[\frac{\partial^2 \ln p\left(\mathbf{p};\bm{\theta}\right)}{\partial \mathbf{t} \partial P_t}\big] \\
                    -\mathbbm{E}\big[\frac{\partial^2 \ln p\left(\mathbf{p};\bm{\theta}\right)}{\partial \mathbf{t} \partial P_t}\big] & -\mathbbm{E}\big[\frac{\partial^2 \ln p\left(\mathbf{p};\bm{\theta}\right)}{\partial P_t^2}\big]
                 \end{bmatrix}
               = \begin{bmatrix}
                    \mathbf{A} & \mathbf{b} \\
                    \mathbf{b}^T & c
                \end{bmatrix},\IEEEeqnarraynumspace
\end{IEEEeqnarray}
where
\begin{IEEEeqnarray}{lll}
\label{Eq:FIM1}
\IEEEyesnumber\IEEEyessubnumber*
\mathbf{A} &=& \sum_{i\in\mathcal{S}}\frac{\mathbf{c}_i\mathbf{c}_i^T}{\sigma_i^2(\ln10)^2\|\mathbf{t}-\mathbf{s}_i\|^4},\\
\mathbf{b} &=& \sum_{i\in\mathcal{S}}\frac{-\mathbf{c}_i}{\sigma_i^2\ln10\|\mathbf{t}-\mathbf{s}_i\|^2},\\
c &=& \sum_{i\in\mathcal{S}}\frac{1}{\sigma_i^2}.
\end{IEEEeqnarray}
For any vector $[\mathbf{x}^T, y]^T \in \mathbb{R}^{k+1}$, the corresponding quadratic form can be written as:
\begin{IEEEeqnarray}{cl}
\label{Eq:FIM_Invert1}
[\mathbf{x}^T,y]\mathbf{F}[\mathbf{x}^T,y]^T = \mathbf{x}^T\mathbf{A}\mathbf{x} + 2y\mathbf{x}^T\mathbf{b} + cy^2.
\end{IEEEeqnarray}
By substituting the expressions of $\mathbf{A}$, $\mathbf{b}$, and $c$ into~\eqref{Eq:FIM_Invert1}, we obtain:
\begin{IEEEeqnarray}{cl}
\label{Eq:FIM_Invert2}
[\mathbf{x}^T,y]\mathbf{F}[\mathbf{x}^T,y]^T = \sum_{i\in\mathcal{S}}\frac{1}{\sigma_i^2}\big(\frac{\mathbf{x}^T\mathbf{c}_i}{\ln10\|\mathbf{t}-\mathbf{s}_i\|^2} + y\big)^2.
\end{IEEEeqnarray}
Therefore, as each term in the summation is nonnegative, $\mathbf{F}$ is positive semidefinite. Moreover, the quadratic form equals to zero if and only if 
\begin{IEEEeqnarray}{cl}
\label{Eq:FIM_Invert3}
\mathbf{x}^T\mathbf{c}_i + y \ln(10)\|\mathbf{t}-\mathbf{s}_i\|^2 = 0, \quad \forall i \in \mathcal{S},
\end{IEEEeqnarray}
which is equivalent to:
\begin{IEEEeqnarray}{cl}
\label{Eq:FIM_Invert4}
[\mathbf{c}_i^T,\ln10\|\mathbf{t}-\mathbf{s}_i\|^2][\mathbf{x}^T,y]^T = 0,\ \forall i\in\mathcal{S}.
\end{IEEEeqnarray}
Let $\mathbf{g}_i = [\mathbf{c}_i^T,\ \ln(10)\|\mathbf{t}-\mathbf{s}_i\|^2]^T$. In practical settings, anchor nodes are typically distributed across the region of interest rather than being aligned along a line. Hence, it is reasonable to assume that the vectors $\mathbf{g}_i$ are linearly independent, which implies that the only vector satisfying~\eqref{Eq:FIM_Invert3} is the zero vector. 
Consequently, the quadratic form in~\eqref{Eq:FIM_Invert1} is strictly positive for every nonzero $[\mathbf{x}^T,y]^T$, and therefore $\mathbf{F}$ is positive definite. This completes the proof.
Accordingly, the root mean square error (RMSE) for estimating $\mathbf{t}$ and $P_t$ is lower bounded by 
\begin{IEEEeqnarray}{llr}
    \label{Eq:CRLB}
    \IEEEyesnumber\IEEEyessubnumber*
    \text{RMSE}_t &=\!\! \sqrt{\|\mathbf{t}-\hat{\mathbf{t}}\|^2} \geq \sqrt{\text{trace}\left([\mathbf{F}^{-1}]_{1:k,1:k}\right)} \triangleq \text{CRLB}_t,\IEEEeqnarraynumspace\\
    \text{RMSE}_p &=\!\! \sqrt{\|P_t-\hat{P}_0\|^2} \geq\nonumber\\
    &\hspace{2cm}\sqrt{\text{trace}\left([\mathbf{F}^{-1}]_{k+1,k+1}\right)} \triangleq \text{CRLB}_p,\IEEEeqnarraynumspace
\end{IEEEeqnarray}
where $\hat{\mathbf{t}}$ and $\hat{P}_0$ are the estimated location and transmit power of the target node, respectively. When $P_t$ is known, the FIM simplifies to $\mathbf{F} = \sum_{i\in\mathcal{S}}\frac{\mathbf{c}_i\mathbf{c}_i^T}{\sigma_i^2(\ln10)^2\|\mathbf{t}-\mathbf{s}_i\|^4}$ and the CRLB of location estimate is given by $\text{CRLB}_t \triangleq \sqrt{\text{trace}\left([\mathbf{F}^{-1}]_{1:k,1:k}\right)}$.

\begin{table*}[!t]
\centering
\caption{Summary of the existing underwater localization techniques.}
\resizebox{\textwidth}{!}{\begin{tabular}{|c|c|cc|c|c|c|c|c|}
\hline
\multirow{2}{*}{Algorithm}                   & \multirow{2}{*}{Unknown parameters} & \multicolumn{2}{c|}{Estimated parameters}                          & \multirow{2}{*}{Method}   & \multirow{2}{*}{Accuracy} & \multirow{2}{*}{Complexity} & \multirow{2}{*}{CRLB derivation} & \multirow{2}{*}{Year} \\ \cline{3-4}
                                             &                                     & \multicolumn{1}{c|}{Location}              & Transmit power        &                           &                           &                             &                                  &                       \\ \hline
SDP-Xu~\cite{SDPXu}    & Transmit power                      & \multicolumn{1}{c|}{$\checkmark$}          & $\times$              & SDP                       & Low                       & Moderate                        & $\times$                         & 2016                  \\ \hline
Inam-Range~\cite{Inam} & None                                & \multicolumn{1}{c|}{$\checkmark$}          & $\times$              & Linear least squares             & Moderate                       & Low                         & $\times$                         & 2019                  \\ \hline
SDSOCP-K~\cite{SDSOCP} & None                                & \multicolumn{1}{c|}{$\checkmark$}          & $\times$              & Mixed SDP-SOCP            & Moderate                  & High                        & $\checkmark$                     & 2019                  \\ \hline
SDSOCP-U~\cite{SDSOCP} & Transmit power                      & \multicolumn{1}{c|}{$\checkmark$}          & $\checkmark$          & Mixed SDP-SOCP            & Moderate                  & High                        & $\checkmark$                     & 2019                  \\ \hline
DL-Salama~\cite{R_AI} & None                                & \multicolumn{1}{c|}{$\checkmark$}          & $\times$              & Deep learning             & High                      & High                        & $\times$                         & 2023                  \\ \hline
OSUL~\cite{CM_S2}     & None                                & \multicolumn{1}{c|}{$\checkmark$}          & $\times$              & Triangulation             & Low                       & Low                         & $\times$                         & 2023                  \\ \hline
RCFLA~\cite{CM_S1}    & None                                & \multicolumn{1}{c|}{$\checkmark$}          & $\times$              & ASM, BFGS                 & Moderate                  & Moderate                    & $\checkmark$                     & 2025                  \\ \hline
\textbf{GUTP}                                & \textbf{Transmit power}             & \multicolumn{1}{c|}{\textbf{$\checkmark$}} & \textbf{$\checkmark$} & \textbf{Bisection method} & \textbf{High}             & \textbf{Low}                & \textbf{$\checkmark$}            & \textbf{2025}         \\ \hline
\end{tabular}}
\label{Tb:ExistingTech}
\end{table*}
\section{Performance analysis}\label{sec:PerformanceAnalysis}
\begin{figure}[!t]
    \centering
    \includegraphics[width=\linewidth]{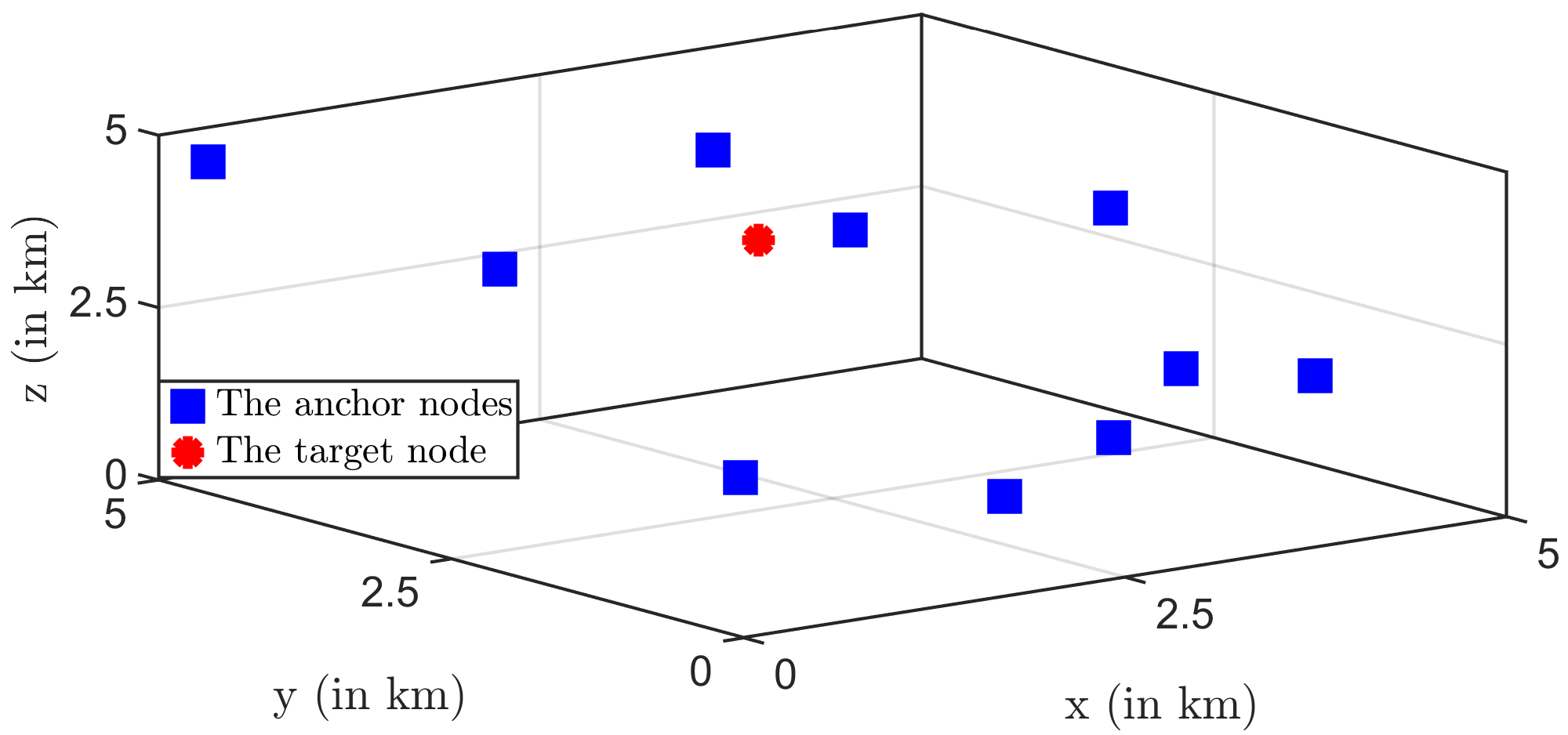}
    \caption{The evaluated UWSN topology.}
    \label{Fig:topology}
\end{figure}

\begin{figure}[!t]
    \centering
    \includegraphics[width=\linewidth]{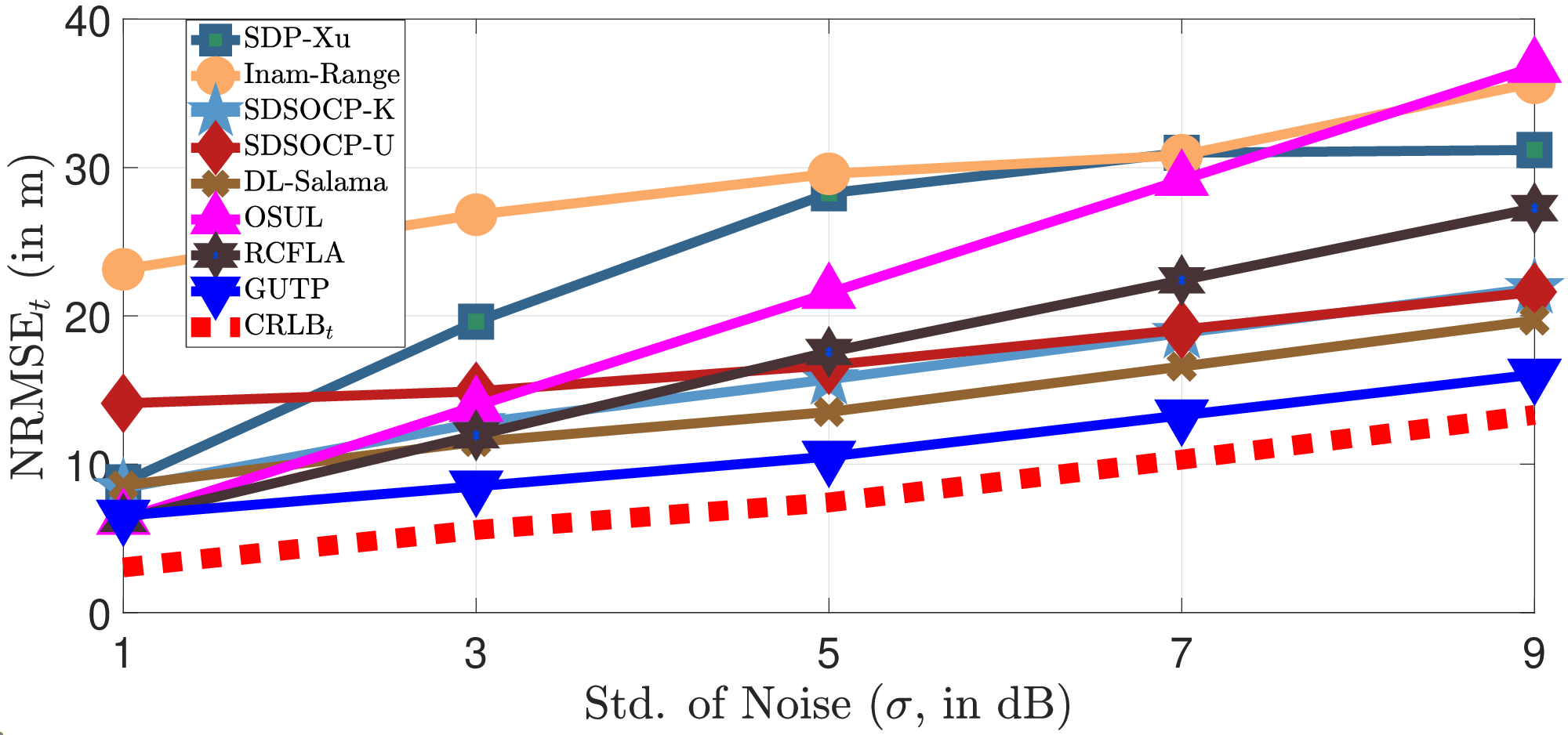}
    \caption{NRMSE of location estimate as a function of $\sigma$.}
    \label{Fig:location_sigma}
\end{figure}

\begin{figure}[!t]
    \centering
    \includegraphics[width=\linewidth]{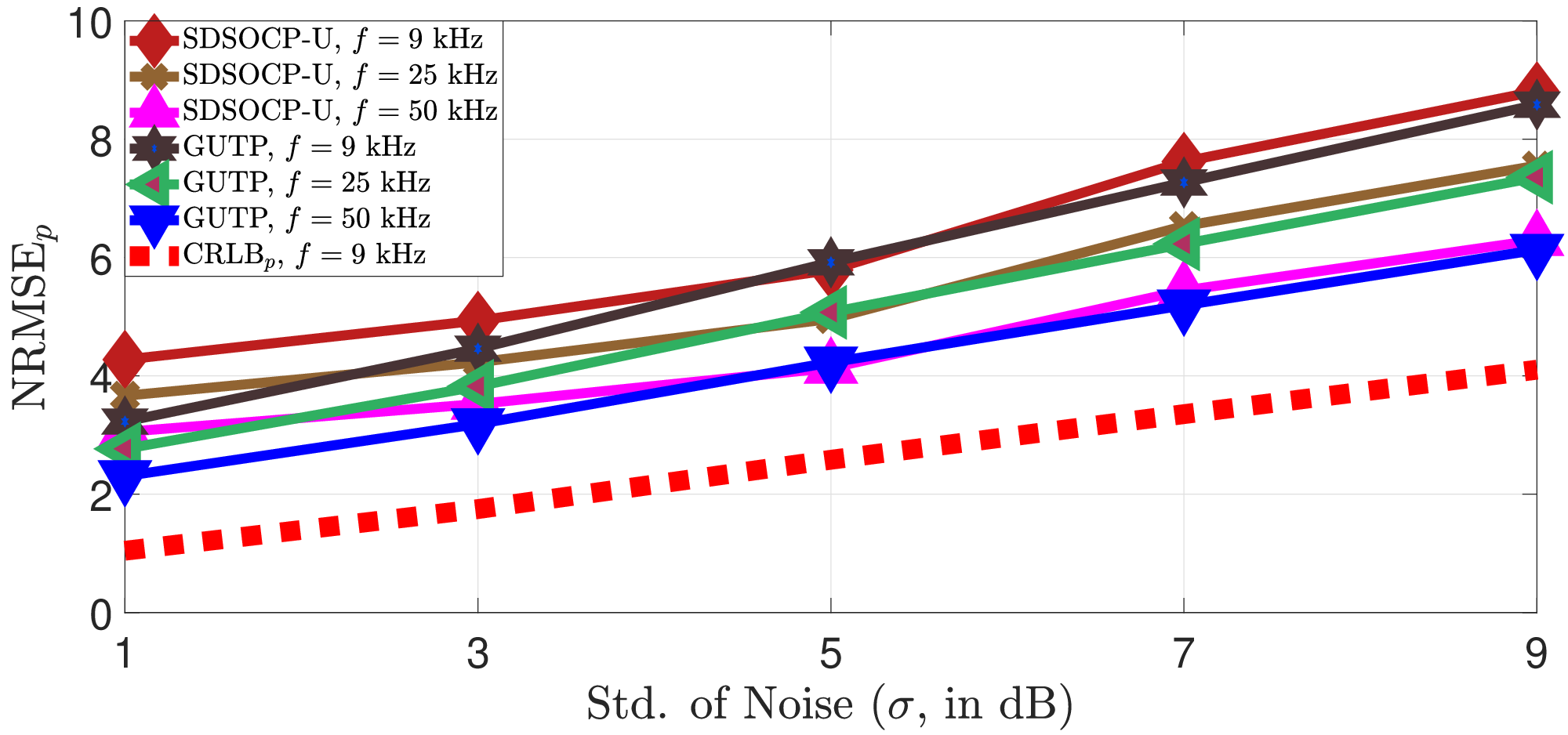}
    \caption{NRMSE of transmit power estimate as a function of $\sigma$.}
    \label{Fig:power_sigma}
\end{figure}

In this section, we perform numerical simulations to evaluate the performance of GUTP in comparison to the localization techniques. To facilitate a fair evaluation of our proposed method, Table~\ref{Tb:ExistingTech} summarizes representative underwater localization techniques reported in the literature and considered in our simulations. This comparison highlights the main features and assumptions of each approach, providing a reference for the subsequent performance analysis. To solve the SDPs associated with these techniques, we use the CVX toolbox with the precision parameter set to $`best"$. In~\cite{Inam}, the authors proposed two localization techniques: one distance-based and one angle-based. Since the RSS-based localization discussed in this paper is a type of distance-based technique, we select the distance-based method from~\cite{Inam} for comparison and refer to it as Inam-Range. Without loss of generality, we assume that $\sigma_i = \sigma$. In underwater communication, acoustic signals below 50~kHz can propagate over 1~km. However, due to challenges in deployment, the number of anchor nodes ($N$) in UWSNs is small~\cite{Review}. Underwater environments are inherently three-dimensional. Accordingly, we consider a $5~\text{km} \times 5~\text{km} \times 5~\text{km}$ network and randomly deploy ten anchor nodes and one target node to emulate a realistic underwater scenario. The considered topology is presented in Fig.~\ref{Fig:topology}, where blue squares denote anchor nodes and the red star denote the target node. Due to variations in experimental settings, the numerical results reported in this paper may differ from those presented in the references summarized in Table~\ref{Tb:ExistingTech}. To facilitate the reproducibility of the simulation results, Table~\ref{Tb:ParaManu} lists the key simulation parameters and the coordinates of all nodes. Throughout the simulations, node positions remain fixed. We assume each anchor node can maintain a stable connection with the target node and that no malicious interference is present. The PLE is set to \( \beta = 2 \), representing spherical propagation, which is an assumption commonly used in previous studies~\cite{CM_S1, CM_S2, SDPXu, SDSOCP}. For the absorption coefficient in~\eqref{Eq:Absorption}, we use a center frequency of \( f = 9~\text{kHz} \), which falls within the typical range of commercial underwater acoustic systems designed for kilometer-scale communication~\cite{Frequency}. To ensure fair comparisons with existing techniques, we evaluate localization performance under noise standard deviations below 10~dB, which aligns with prior work~\cite{SDSOCP, CM_S1}. We consider the normalized root mean square error (NRMSE) as the performance metric. The NRMSE of location and transmit power estimate is given by $\text{NRMSE}(\hat{\mathbf{t}}) = \sqrt{\frac{1}{M_c}\sum\nolimits_{m=1}^{M_c}\|\mathbf{t}-\hat{\mathbf{t}}^m\|^2}$, $\text{NRMSE}(\hat{P}_0) = \sqrt{\frac{1}{M_c}\sum\nolimits_{m=1}^{M_c}\|P_t-\hat{P}_0^m\|^2}$, where $\hat{\mathbf{t}}^m$ and $\hat{P}_0^m$ denote the estimated location and transmit power of the target node in the $m^{\text{th}}$ Monte Carlo simulation. $M_c$ represents the number of Monte Carlo simulations. In this paper, we obtain all the simulation results from 3000 Monte Carlo simulations. In this section, all localization techniques summarized in Table~\ref{Tb:ExistingTech} are evaluated under a uniform network configuration, with identical node layouts and numbers of nodes. The corresponding RMSE results across different scenarios are reported in Fig.~\ref{Fig:location_sigma}--Fig.~\ref{Fig:R1C2_sensitivity} and are further analyzed in Section~\ref{subsec:location}--Section~\ref{subsec:sensitivity}.
\begin{table}
\centering
\caption{Simulation parameters}
\label{Tb:ParaManu}
\resizebox{.48\textwidth}{!}{\begin{tabular}{|c|c|c|}
\hline
\textbf{Parameters} & \textbf{Description}                                                                            & \textbf{Value}                                                                                                                                         \\ \hline
$\mathbf{s}_i$      & The location of the anchor nodes (in km)                                                         & \begin{tabular}[c]{@{}c@{}}$[3.38,1.27,4.46]^T$, $[4.22,0.62,2.03]^T$, \\ $[4.29,1.87,1.54]^T$, $[0.03,0.08,2.27]^T$, \\ $[0.44,2.65,3.97]^T$, $[2.25,3.19,4.82]^T$, \\ $[0.25,4.91,4.57]^T$, $[2.59,0.22,1.89]^T$, \\ $[3.40,3.52,3.11]^T$, and $[2.87,1.52,0.35]^T$\end{tabular} \\ \hline
$\mathbf{t}$        & The location of the target node (in km)                                                          & $[2.98,3.75,3.00]^T$\\ \hline
$\beta$~\cite{CM_S1, CM_S2}              & The PLE                                                                                         & 2                                                                                                                                                      \\ \hline
$f$~\cite{Frequency}                 & \begin{tabular}[c]{@{}c@{}}The center frequency of \\ the acoustic signal (in kHz)\end{tabular} & 9                                                                                                                                                      \\ \hline
$\alpha$            & The absorption coefficient (in dB/m)                                                            & $9.86\times10^{-4}$                                                                                                                                    \\ \hline
$\sigma$~\cite{SDSOCP, CM_S1}            & The noise standard deviation (in dB)                                                            & $[1,3,5,7,9]^T$                                                                                                                                        \\ \hline
$M_c$               & The number of Monte Carlo simulations                                                           & 3000                                                                                                                                                   \\ \hline
\end{tabular}}
\end{table}
\subsection{Performance in Estimating the Location}\label{subsec:location}
In practice, the variance of ambient noise depends on the surrounding ocean environment.  In deep-sea regions far from human activities, the background is typically quiet, as the major natural sources of noise (e.g., wind, waves, and rain) are often negligible at the operating frequencies of acoustic signals. Conversely, in coastal areas or shipping lanes, anthropogenic sources and natrual sources, such as industrial activities and rains can significantly increase the noise level. Therefore, in this study, we evaluate localization performance under different noise standard deviation to reflect these practical scenarios. In Fig.~\ref{Fig:location_sigma}, we examine the performance of localization techniques in estimating the target node's location as a function of the noise variance. Fig.~\ref{Fig:location_sigma} presents the superior performance of GUTP relative to existing localization techniques across the evaluated scenarios. SDSOCP-K~\cite{SDSOCP} outperforms SDSOCP-U by exploiting knowledge of the transmit power. However, SDSOCP-K omits the Gaussian randomization step after solving the SDP, which leads to accuracy degradation due to the relaxed rank constraint—an effect that becomes more pronounced in high-noise conditions. In contrast, GUTP achieves performance closest to the $\text{CRLB}_t$, as it avoids the accuracy loss associated with SDR by directly solving the proposed GTRS in~\eqref{Eq:estimator_np}. This enables GUTP to maintain reliable localization accuracy even under challenging noise conditions, making it more robust and practical for underwater environments. DL-Salama integrates deep learning with Kalman filtering by leveraging RSS measurements to train a neural network and applying an iterative algorithm to estimate the target node’s position. Nevertheless, its performance is surpassed by GUTP, since GUTP is formulated as a GTRS (refer to~\eqref{Eq:estimator_np}), for which convergence to the global minimum is theoretically guaranteed~\cite{Bisection,GTRS2}. Furthermore, DL-Salama assumes prior knowledge of the target node’s transmit power, a limitation in realistic scenarios where such information is rarely available. In addition, the training of neural networks and the iterative process of Bayesian optimization incur high computational complexity (refer to Table~\ref{Tb:complexity}), reducing its suitability for underwater applications with constrained computational resources. OSUL estimates distances between the target and anchor nodes using RSS measurements and then applies a triangulation variant to solve the resulting linear system. Under the assumption of known transmit power, OSUL achieves better performance than SDP-Xu when $\sigma = 1$~dB. However, its accuracy deteriorates significantly as the noise variance increases, owing to poor distance estimation in high-noise conditions. RCFLA employs an iterative BFGS trust-region method and achieves moderate localization performance. Its effectiveness, however, strongly depends on the accuracy of the initial estimate provided by the ASM algorithm. A large deviation from the true value often results in convergence to a local minimum. Moreover, RCFLA relies on approximate prior knowledge of the transmit power, further limiting its practicality compared to GUTP, which operates effectively without requiring such information.

\subsection{Performance in Estimating the Transmit Power}\label{subsec:power}
This study examines the effectiveness of localization techniques in estimating the target node's transmit power. Fig.~\ref{Fig:power_sigma} exhibits the NRMSE of the transmit power estimates for the target node as the noise standard deviation $\sigma$ increases from 1 to 9~dB, under center frequencies of 9~kHz, 25~kHz, and 50~kHz. It can be observed that GUTP consistently outperforms SDSOCP-U at all frequencies. The inferior performance of SDSOCP-U arises from the relaxation of the rank constraint in the SDR process, which introduces accuracy loss. As $\sigma$ increases, the performance gap between GUTP and SDSOCP-U becomes more pronounced. This is because SDSOCP-U estimates the transmit power based on the relationship between the link distance and the target node’s power, where the distance estimation is highly sensitive to noise variance (refer to Fig.~\ref{Fig:location_sigma}). Consequently, errors in location estimation propagate to the transmit power estimates, degrading performance. Furthermore, Fig.~\ref{Fig:power_sigma} demonstrates that higher center frequencies improve the accuracy of localization techniques. At elevated frequencies, signal attenuation with distance becomes more pronounced, enabling RSS measurements to capture finer variations in the target node’s position. This increased resolution leads to more accurate distance estimation, thereby reducing the NRMSE of the transmit power estimates.

\begin{figure}
    \centering
    \includegraphics[width=\linewidth]{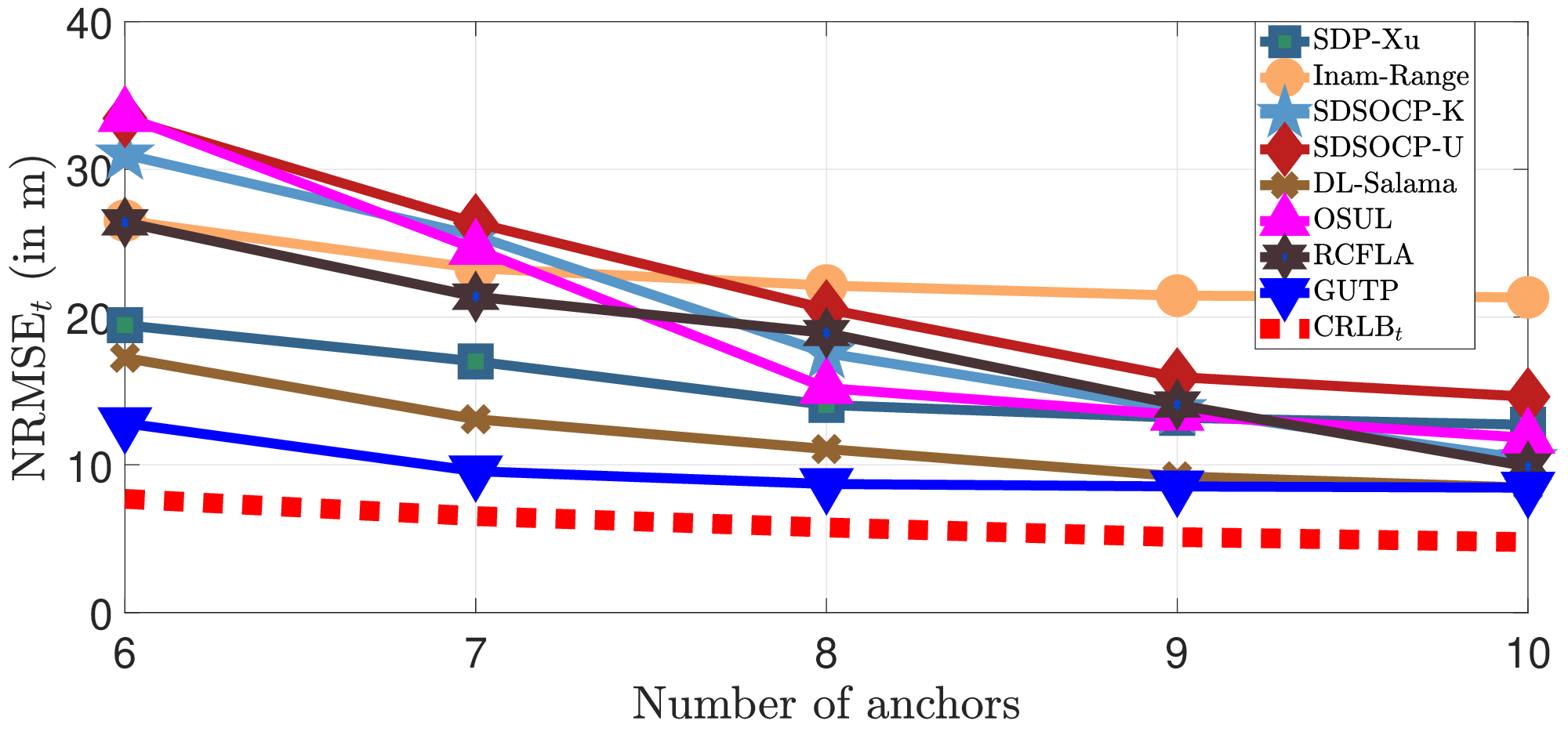}
    \caption{The effect of the number of anchor nodes on the performance of the localization techniques.}
    \label{Fig:location_Na}
\end{figure}
\begin{figure}[!t]
    \centering
    \subfigure[The effect of the PLE on the performance of the localization techniques.]{
    \begin{minipage}[t]{\linewidth}
    \centering
    \includegraphics[width=\linewidth]{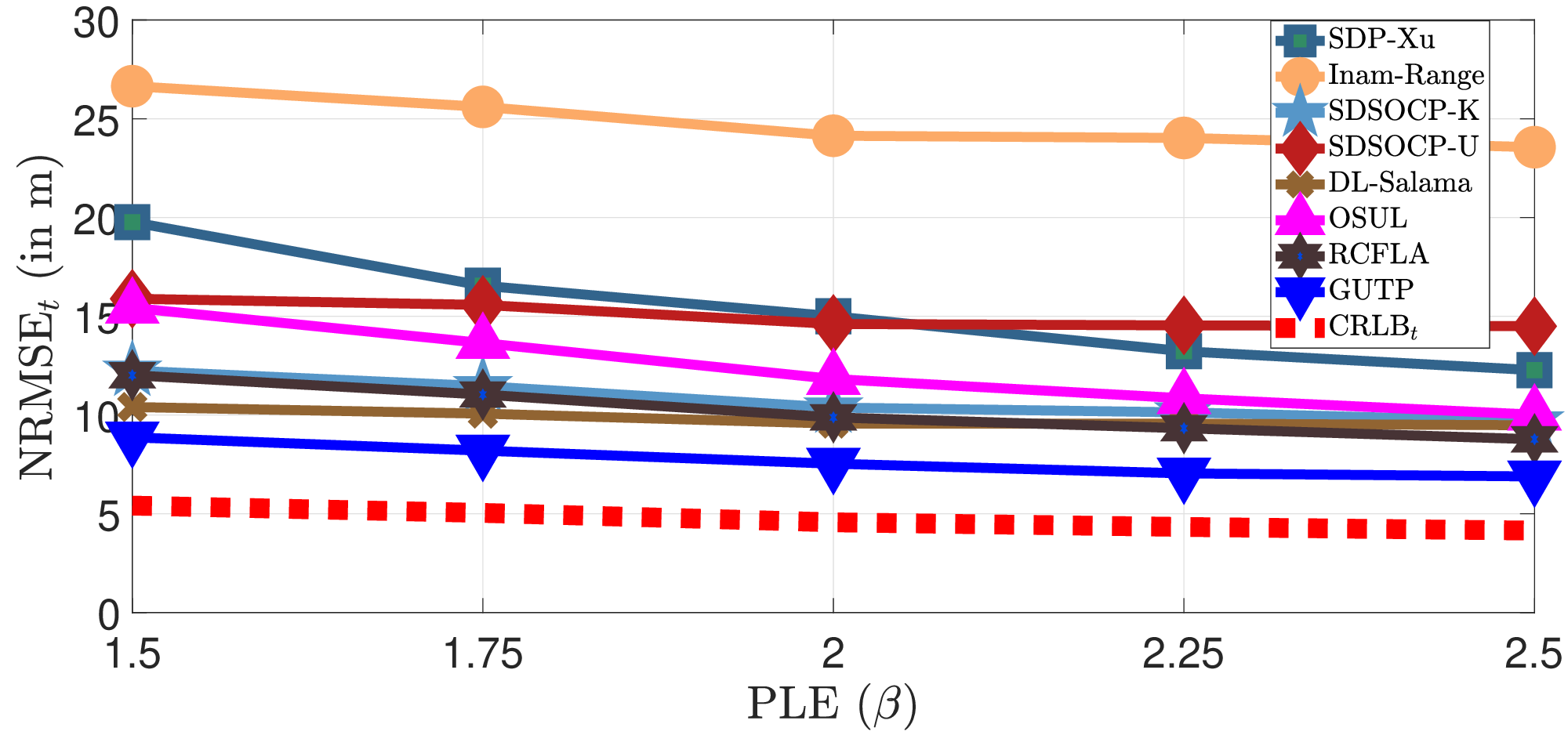}
    \label{Fig:location_beta}
    \end{minipage}
    }
    \subfigure[The effect of the signal frequency on the performance of the localization techniques.]{
    \begin{minipage}[t]{\linewidth}
    \centering
    \includegraphics[width=\linewidth]{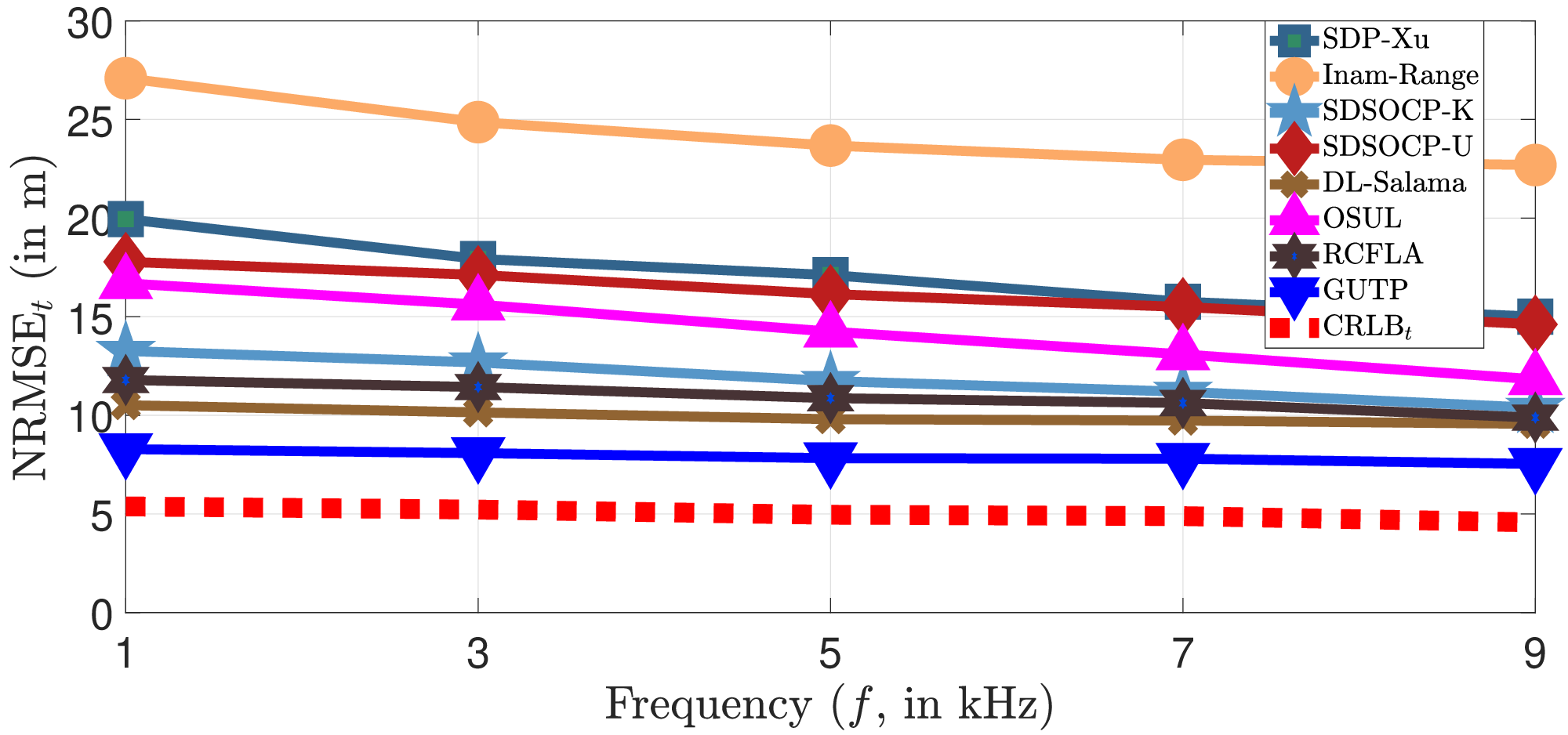}
    \label{Fig:location_alpha}
    \end{minipage}
    }
    \caption{The effect of different PLEs and signal frequencies on the performance of localization techniques.}
    \label{Fig:RMSE_location_BetaAlpha}
\end{figure}
\subsection{Effect of the Number of Anchor Nodes}\label{subsec:Na}
This study examines the performance of various localization techniques as the number of anchor nodes \(N\) in Fig.~\ref{Fig:topology} increases from 6 to 10. Fig.~\ref{Fig:location_Na} presents the NRMSE for estimating the target node’s location, with the noise standard deviation \(\sigma\) fixed at 2~dB. As the number of anchor nodes $N$ increases, the accuracy of localization techniques improves owing to the availability of additional target–anchor links, which provide more information for position estimation. Fig.~\ref{Fig:location_Na} demonstrates that GUTP consistently outperforms existing techniques across all scenarios, particularly when $N < 8$, due to its theoretical guarantee of convergence to the global minimum. RCFLA achieves accuracy comparable to GUTP when $N > 8$; however, for smaller $N$, the initial estimate generated by ASM becomes unreliable, causing RCFLA to converge to a local minimum and thereby reducing accuracy. SDSOCP-K exhibits localization accuracy similar to GUTP but requires prior knowledge of the transmit power. Moreover, SDSOCP-K relies on computationally intensive SDP and SOCP formulations, whereas GUTP achieves comparable or superior accuracy using the more efficient bisection method (refer to Table~\ref{Tb:complexity}). By contrast, OSUL benefits less from an increasing number of anchor nodes. Its performance is more sensitive to noise, leading to significant degradation as the noise variance grows (refer to Fig.~\ref{Fig:location_sigma}). Finally, as shown in Fig.~\ref{Fig:location_Na}, the accuracy gain becomes marginal when $N > 8$, since the additional anchor nodes provide limited incremental information beyond this threshold.

\begin{figure}[!t]
    \centering
    \subfigure[NRMSE of location estimate as a function of $\sigma$ for various noise scenarios.]{
    \begin{minipage}[t]{\linewidth}
    \centering
    \includegraphics[width=\linewidth]{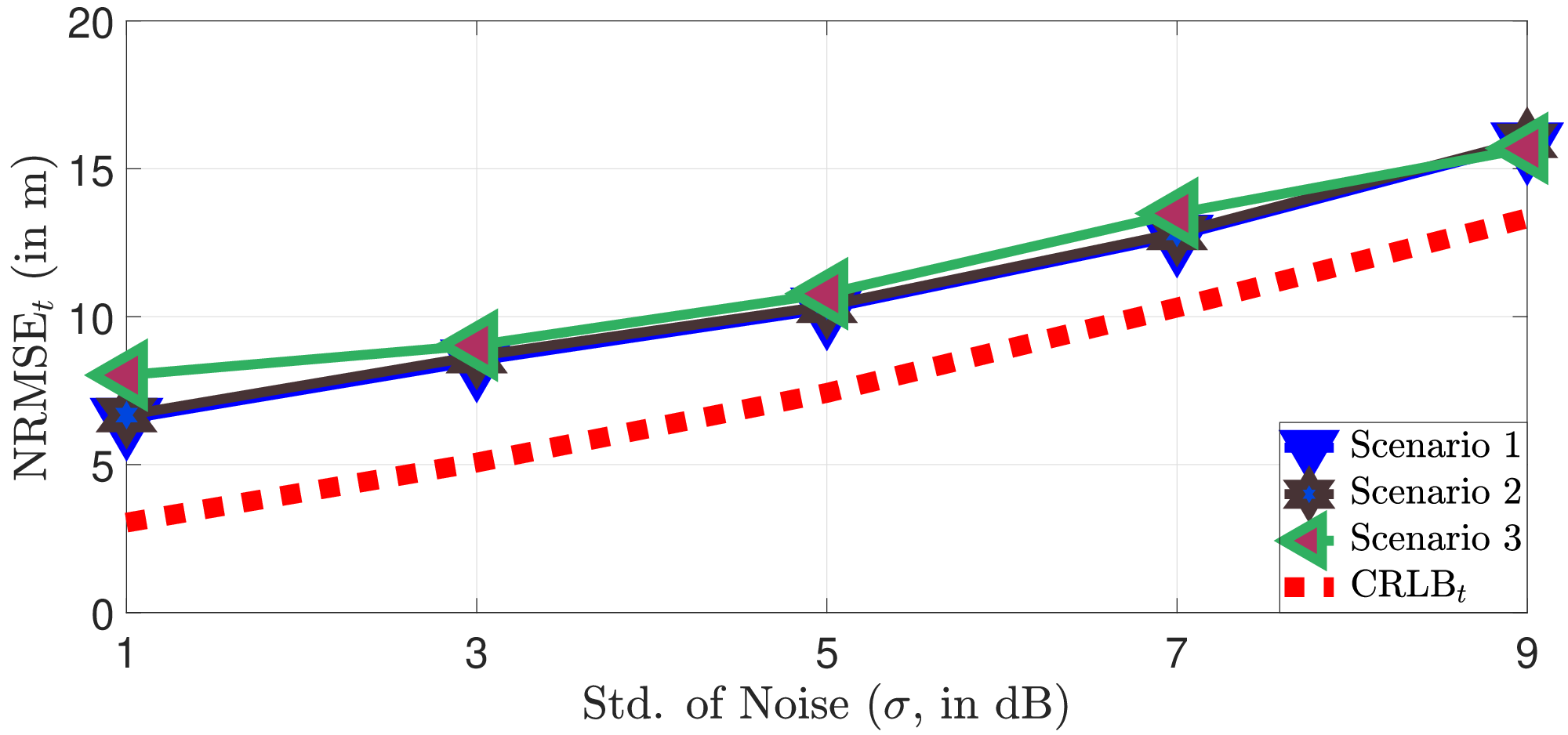}
    \label{Fig:R1C1_noise_location}
    \end{minipage}
    }
    \subfigure[NRMSE of transmit power estimate as a function of $\sigma$ for various noise scenarios.]{
    \begin{minipage}[t]{\linewidth}
    \centering
    \includegraphics[width=\linewidth]{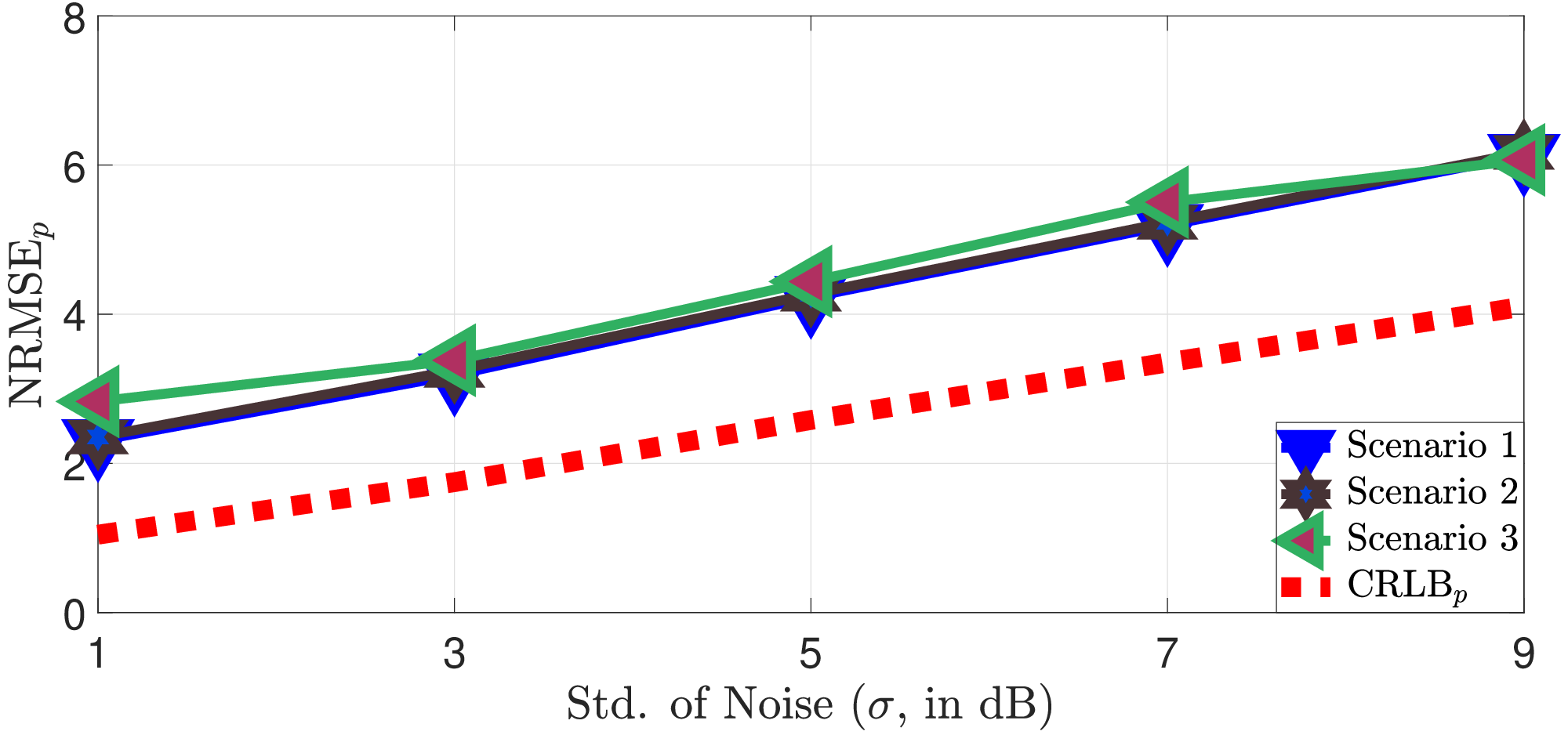}
    \label{Fig:R1C1_noise_power}
    \end{minipage}
    }
    \caption{The effect of different noise conditions on the performance of GUTP.}
    \label{Fig:R1C1_noise}
\end{figure}
\subsection{Effect of the PLE and absorption coefficient}\label{subsec:BetaAlpha}
\begin{figure}[!t]
    \centering
    \subfigure[Sensitivity of GUTP's location estimates to bias in model parameters.]{
    \begin{minipage}[t]{\linewidth}
    \centering
    \includegraphics[width=\linewidth]{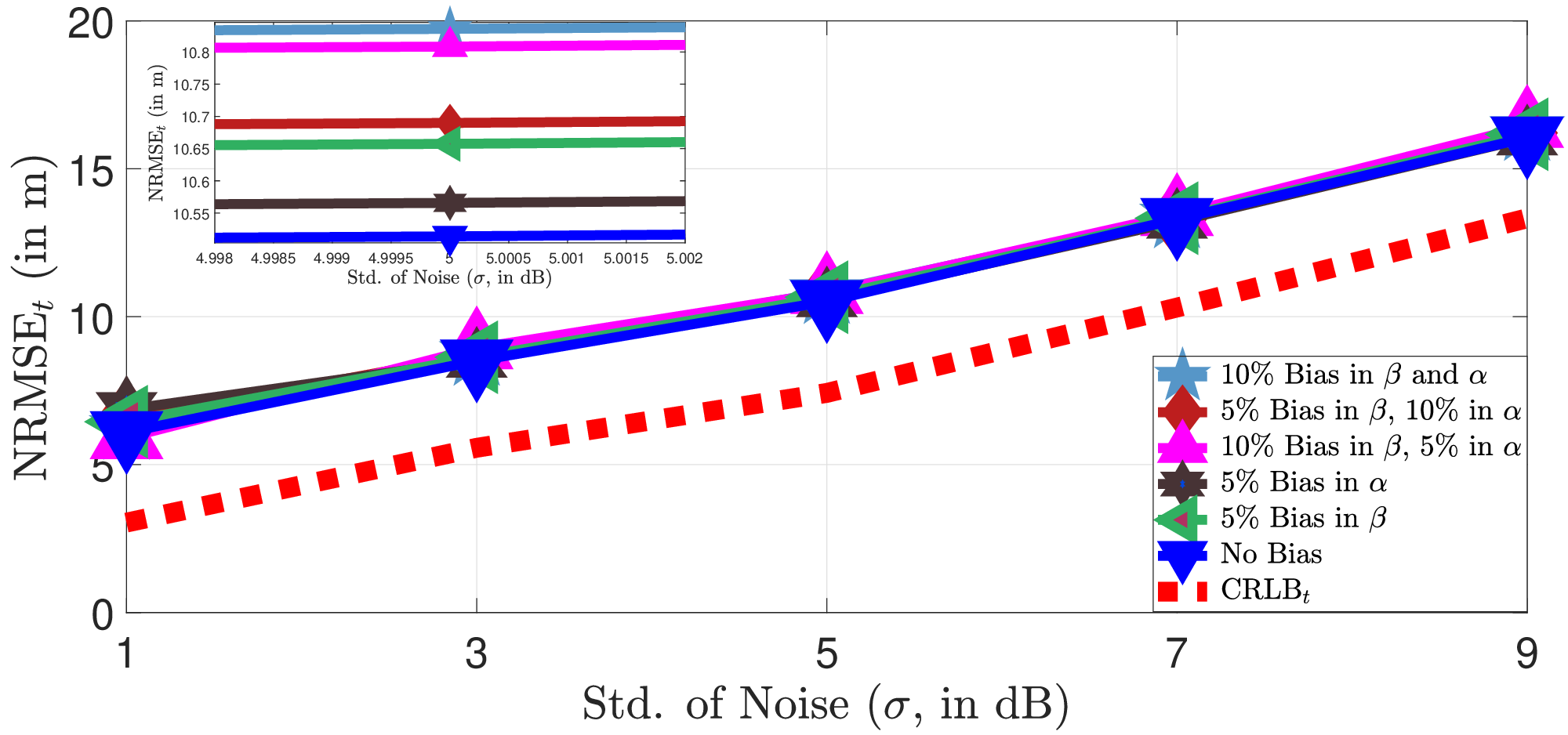}
    \label{Fig:R1C2_sensitivity_location}
    \end{minipage}
    }
    \subfigure[Sensitivity of GUTP's transmit power estimates to bias in model parameters.]{
    \begin{minipage}[t]{\linewidth}
    \centering
    \includegraphics[width=\linewidth]{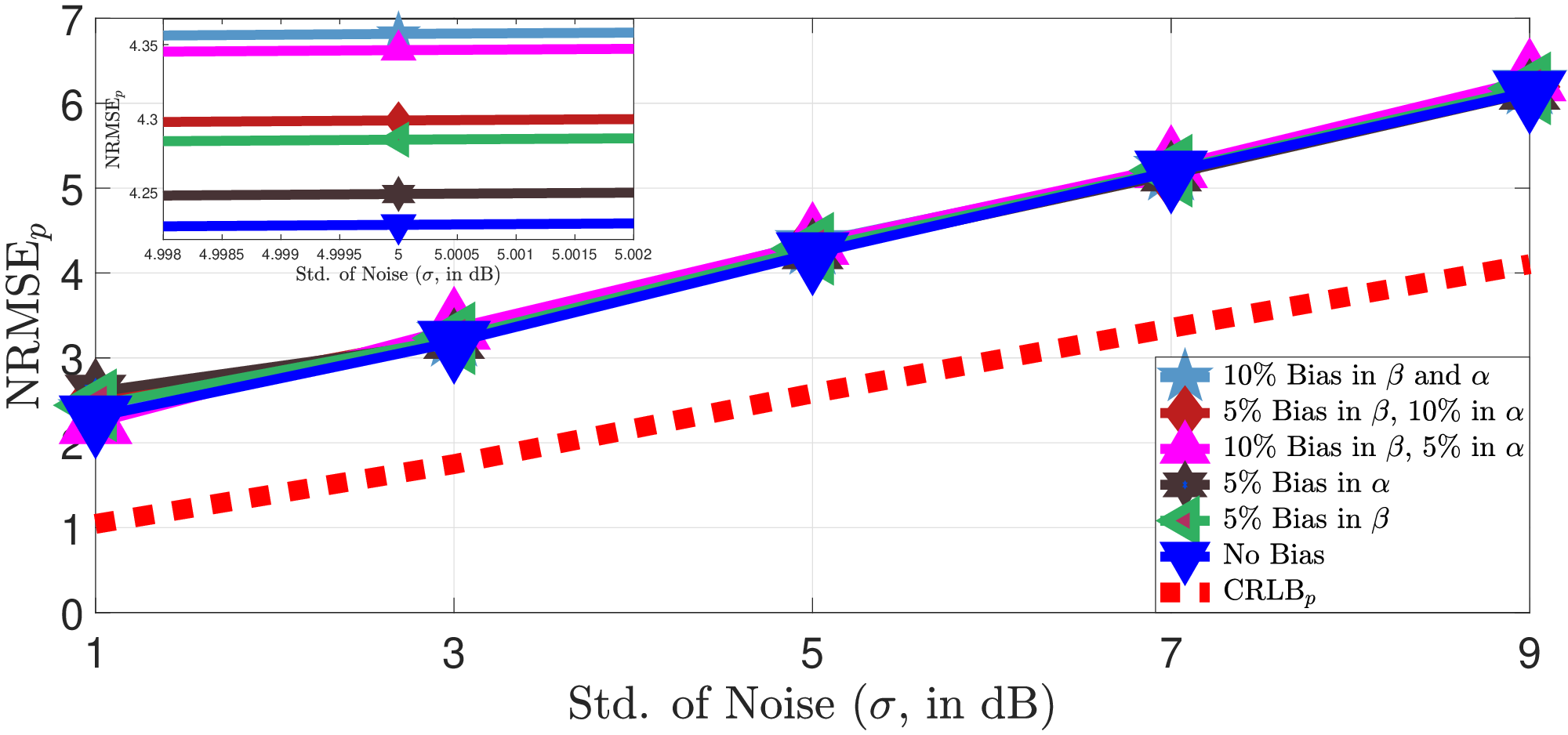}
    \label{Fig:R1C2_sensitivity_power}
    \end{minipage}
    }
    \caption{Sensitivity analysis of GUTP with respect to bias in model parameters.}
    \label{Fig:R1C2_sensitivity}
\end{figure}

% Fig~\ref{Fig:location_beta}, and Fig~\ref{Fig:location_alpha})
In this study, we investigate the impact of the PLE and the absorption coefficient on the performance of localization techniques. Fig.~\ref{Fig:location_beta} displays the NRMSE of the target node’s location estimation as the PLE \(\beta\) increases from 1.5 to 2.5. In practice, \(\beta < 2\) corresponds to propagation conditions typically encountered in shallow water environments. As \(\beta\) increases, the localization accuracy improves owing to the enhanced distance resolution resulting from greater attenuation. Fig.~\ref{Fig:location_alpha} presents the effect of the absorption coefficient (according to~\eqref{Eq:Absorption}, the absorption coefficient $\alpha$ exhibits a monotonic increase with increasing signal frequency), showing that GUTP consistently outperforms existing techniques across all evaluated scenarios. The accuracy of localization improves with increasing signal frequency, since higher frequencies correspond to higher absorption coefficients. However, the performance gain attributable to absorption is smaller than that achieved by increasing the PLE, suggesting that the PLE exhibits a more significant influence on localization performance. This observation is consistent with the sensitivity analysis presented in Fig.~\ref{Fig:R1C2_sensitivity} of Section~\ref{subsec:sensitivity}.

\subsection{Effect of Noise Characteristics on Localization Performance}\label{subsec:R1C1_noise}
In the derivation of GUTP, the measurement noise term $n_i$ is assumed to follow a Gaussian distribution with zero mean. This modeling reflects the common assumption in signal processing that measurement errors are caused by random fluctuations and symmetrically distributed around zero~\cite{FCUP, SDP_Zou, SDP_Wang}. However, in real underwater environments, the statistical characteristics of noise can be more complex, including non-zero-mean Gaussian noise and impulsive noise~\cite{urick1975principles}. In this section, we examine the impact of different noise characteristics on the performance of GUTP proposed in~\eqref{Eq:estimator_np}. Specifically, we consider the following three scenarios: 
\begin{itemize}
    \item Scenario 1: \( n_i \) is a zero-mean Gaussian random variable, consistent with the model adopted in the manuscript;
    \item Scenario 2: \( n_i \) is a Gaussian random variable with a mean of 2~dB;
    \item Scenario 3: \( n_i \) is a linear combination of a Gaussian random variable with a mean of 2~dB and an impulsive noise, both with the same standard deviation.
\end{itemize}
The impulsive noise is simulated using a random variable uniformly distributed over the interval \([0, \alpha]\)~dB, where $\alpha$ is a positive scalar. The value of $\alpha$ is carefully chosen to ensure that the resulting noise standard deviation is consistent across all three evaluated scenarios, thereby enabling a fair comparison of localization performance under different noise characteristics. Fig.~\ref{Fig:R1C1_noise} presents the NRMSE of the estimated location and transmit power as a function of the noise standard deviation. As shown in Fig.~\ref{Fig:R1C1_noise_location}, GUTP achieves the best performance in Scenario~1, where $n_i$ follows a zero-mean Gaussian distribution without impulsive noise. Its performance degrades when $n_i$ has a non-zero mean and deteriorates further in Scenario~3, where $n_i$ contains both non-zero-mean Gaussian noise and impulsive noise. This degradation occurs because the LS-based estimator in~\eqref{Eq:np7} becomes biased under non-zero-mean noise, thereby reducing localization accuracy. A similar trend is observed in Fig.~\ref{Fig:R1C1_noise_power} for transmit power estimation. To sum up, Fig.~\ref{Fig:R1C1_noise} demonstrates that noise characteristics significantly influence localization performance. Nonetheless, the performance variation of GUTP across different noise conditions is relatively small, underscoring its robustness. Furthermore, since the statistical properties of noise are generally unknown in practical underwater environments, maximum likelihood estimation becomes difficult to apply. The ability of GUTP to maintain strong performance regardless of the underlying noise distribution highlights its practical applicability in real-world underwater localization scenarios.

\subsection{Sensitivity Analysis to Bias in Prior Knowledge of Model Parameters}\label{subsec:sensitivity}
The channel model in~\eqref{Eq:CM} reveals that the RSS measurement depends not only on the transmit power \( P_t \) and the distance \( d_i \), but also on the PLE (\( \beta \)) and the absorption coefficient (\( \alpha \)), which we refer to as model parameters in this section. In practical applications, prior knowledge of these model parameters can be obtained through channel estimation techniques~\cite{Underwater_ChannelEstimate1, Underwater_ChannelEstimate2}. However, such prior knowledge is usually biased with respect to the true values due to environmental variability or estimation errors~\cite{CM_S1}. Therefore, it is essential to examine the sensitivity of estimation accuracy to biases in model parameters in order to evaluate the robustness of localization techniques. In this study, we conduct the sensitivity analysis of GUTP to biased prior knowledge of model parameters by evaluating its location estimation accuracy under six scenarios: (a) no bias; (b) a 5\% bias in (\( \beta \)); (c) a 5\% bias in (\( \alpha \)); (d) a 10\% bias in \( \beta \) and a 5\% bias in \( \alpha \); (e) a 10\% bias in \( \alpha \) and a 5\% bias in \( \beta \); and (f) a 10\% bias in both \( \beta \) and \( \alpha \). Fig.~\ref{Fig:R1C2_sensitivity} presents the performance of GUTP for each scenario. As shown in Fig.~\ref{Fig:R1C2_sensitivity_location}, GUTP attains the highest localization accuracy when the model parameters are unbiased, corresponding to perfect prior knowledge. Among the single-parameter bias scenarios, a 5\% bias in $\alpha$ results in smaller error compared to a 5\% bias in $\beta$, indicating that localization accuracy is more sensitive to the PLE than to the absorption coefficient. The lowest performance is observed under simultaneous 10\% biases in both parameters; however, the overall differences in NRMSE across the six scenarios remain marginal. These results demonstrate that GUTP is highly robust to imperfect prior knowledge of model parameters, underscoring its practicality for real-world underwater localization. Similarly, Fig.~\ref{Fig:R1C2_sensitivity_power} shows that the sensitivity of GUTP’s transmit power estimation to biased model parameters closely follows the trends observed for localization. The unbiased case again achieves the lowest NRMSE, while GUTP exhibits a slightly greater sensitivity to biases in the PLE than in the absorption coefficient, consistent with the results in Fig.~\ref{Fig:R1C2_sensitivity_location}. Despite these variations, the overall degradation remains minimal, further confirming the robustness of GUTP in scenarios where prior knowledge of model parameters is inherently imperfect.

\begin{table}[!t]
\centering
\caption{Complexity of the localization techniques.}
\label{Tb:complexity}
\resizebox{.48\textwidth}{!}{\begin{tabular}{|c|c|c|c|}
\hline
Algorithms & Computational complexity & CPU runtime (in s)   & Year     \\ \hline
SDP-Xu~\cite{SDPXu}                        & $\mathcal{O}\left(N^{3.5}\right)$            & 1.24 & 2016 \\ \hline
Inam-Range~\cite{Inam}                     & $\mathcal{O}\left(N\right)$            & 1.46 & 2019 \\ \hline
SDSOCP-K~\cite{SDSOCP}                     & $\mathcal{O}\left((N+5)^{3.5}\right)$        & 2.83 & 2019 \\ \hline
SDSOCP-U~\cite{SDSOCP}                     & $\mathcal{O}\left((N+6)^{3.5}\right)$        & 2.94 & 2019\\ \hline
DL-Salama~\cite{R_AI}                     & $\mathcal{O}\left(N_bN_eN_sN_lN_n\right)$        & 18.25 & 2023 \\ \hline
OSUL~\cite{CM_S2}                     & $\mathcal{O}\left(N\right)$        & 0.27 & 2023\\ \hline
RCFLA~\cite{CM_S1}                     & $\mathcal{O}\left(2N+\tau\right)$        & 4.84 & 2025\\ \hline
\textbf{GUTP}                              & $\mathbf{\mathcal{O}\left(i^{\text{itr}}N\right)}$    & \textbf{0.09} & 2025\\ \hline
\end{tabular}}
\end{table}
\subsection{Complexity Analysis}\label{subsec:complexity}
In this section, we evaluate the worst-case computational complexity of the localization techniques discussed in Section~\ref{sec:PerformanceAnalysis}. The computational complexity and CPU runtime for one Monte Carlo simulation are presented in Table~\ref{Tb:complexity}, where ${i^{\text{itr}}}$ denotes the number of iterations in the bisection method for GUTP. In Table~\ref{Tb:complexity}, $N_b$, $N_e$, $N_s$, $N_l$, and $N_n$ denote the number of Bayesian optimization iterations, the number of epochs per training session, the number of samples in the training dataset, the number of layers in the neural network, and the number of neurons per layer, respectively. The parameter $\tau$ in RCFLA represents the maximum number of iterations of the BFGS trust-region method. In Table~\ref{Tb:complexity}, OSUL estimates the target node’s location by solving a system of linear equations, which yields low computational complexity. However, its localization accuracy is inferior to that of GUTP (refer to Fig.~\ref{Fig:location_sigma} and Fig.~\ref{Fig:location_Na}). Moreover, OSUL assumes prior knowledge of the transmit power, which limits its practicality compared to GUTP. DL-Salama leverages neural networks and iterative Bayesian optimization to improve localization accuracy, but this comes at the cost of substantial computational complexity. Although online training can partially mitigate this issue, the sparse node deployment in UWSNs makes it difficult to collect sufficient training data, thereby degrading the performance of DL-Salama. Compared to RCFLA, GUTP exhibits higher worst-case computational complexity but achieves shorter CPU runtime per Monte Carlo simulation. This efficiency arises from the use of the bisection method, which converges more rapidly than the BFGS trust-region method employed in RCFLA. Consequently, GUTP demonstrates superior computational efficiency in practical scenarios. In addition, GUTP maintains linear complexity with respect to $N$, ensuring strong scalability in dense networks. For CPU runtime, SDSOCP-U requires slightly more time than SDSOCP-K due to the additional computation for transmit power estimation following the SDP step. In contrast, GUTP completes localization in 90~ms, which is 14$\times$ faster than SDP-XU, underscoring the efficiency gains brought by the bisection method. Overall, extensive numerical simulations confirm that GUTP simultaneously achieves higher accuracy and lower complexity than existing localization techniques, thereby validating its effectiveness in UWSNs. It is noteworthy that GUTP achieves strong performance in the evaluated scenarios, however, several practical limitations can arise in real-world underwater deployments. Environmental factors such as time-varying attenuation, scattering, and perturbed anchors are hard to characterize accurately and introduce additional errors beyond those reflected in our simulations. A detailed examination of these complexities, along with their associated performance analysis and validation, lies beyond the present scope and forms a direction for subsequent research.

\section{Conclusion}\label{sec:Conclusion}
In this paper, we have addressed the localization problem in UWSNs where the transmit power of the target node is unknown. We have proposed a localization technique (GUTP) that assigns distance-based weights to RSS measurements, providing more significance to measurements from closer anchor nodes. Utilizing these weighted RSS measurements, we have developed an estimator based on the GTRS framework. We have solved the estimator using a simple bisection method to jointly estimate both the target node's location and its transmit power. Additionally, we have derived the CRLBs for RSS-based underwater localization under both scenarios with known and unknown transmit power. Through extensive simulations, we have demonstrated that GUTP outperforms existing localization techniques in terms of both localization accuracy and computational efficiency. Our future work will focus on developing localization techniques capable of operating in the presence of malicious nodes. The direction for future work is to extend the proposed framework to address network-level challenges such as dead nodes and void regions. Extending the modeling framework to incorporate additional environmental factors is the subject of ongoing work and will be reported in a subsequent study. Ocean experiments are to be undertaken as appropriate experimental conditions permit.

\bibliography{IEEEabrv,ref}

\begin{thebibliography}{10}
\providecommand{\url}[1]{#1}
\csname url@samestyle\endcsname
\providecommand{\newblock}{\relax}
\providecommand{\bibinfo}[2]{#2}
\providecommand{\BIBentrySTDinterwordspacing}{\spaceskip=0pt\relax}
\providecommand{\BIBentryALTinterwordstretchfactor}{4}
\providecommand{\BIBentryALTinterwordspacing}{\spaceskip=\fontdimen2\font plus
\BIBentryALTinterwordstretchfactor\fontdimen3\font minus \fontdimen4\font\relax}
\providecommand{\BIBforeignlanguage}[2]{{%
\expandafter\ifx\csname l@#1\endcsname\relax
\typeout{** WARNING: IEEEtran.bst: No hyphenation pattern has been}%
\typeout{** loaded for the language `#1'. Using the pattern for}%
\typeout{** the default language instead.}%
\else
\language=\csname l@#1\endcsname
\fi
#2}}
\providecommand{\BIBdecl}{\relax}
\BIBdecl

\bibitem{Review}
J.~Luo, Y.~Yang, Z.~Wang, and Y.~Chen, ``Localization algorithm for underwater sensor network: A review,'' \emph{IEEE Internet Things J.}, vol.~8, no.~17, pp. 13\,126--13\,144, Sep. 2021.

\bibitem{Xue2}
X.~Zhang, Z.~Zheng, W.-Q. Wang, and H.~C. So, ``{DOA} estimation of mixed circular and noncircular sources using nonuniform linear array,'' \emph{IEEE Trans. Aerosp. Electron. Syst.}, vol.~58, no.~6, pp. 5703--5710, Dec. 2022.

\bibitem{Inam}
I.~Ullah, J.~Chen, X.~Su, C.~Esposito, and C.~Choi, ``Localization and detection of targets in underwater wireless sensor using distance and angle based algorithms,'' \emph{IEEE Access}, vol.~7, pp. 45\,693--45\,704, Apr. 2019.

\bibitem{Intro_underwater1}
J.~Luo, Y.~Chen, M.~Wu, and Y.~Yang, ``A survey of routing protocols for underwater wireless sensor networks,'' \emph{IEEE Commun. Surv. Tutor.}, vol.~23, no.~1, pp. 137--160, Jan. 2021.

\bibitem{Intro_underwater2}
J.~Zhou, Y.~Wang, C.~Li, and W.~Zhang, ``Multicolor light attenuation modeling for underwater image restoration,'' \emph{IEEE J. Ocean. Eng.}, vol.~48, no.~4, pp. 1322--1337, Oct. 2023.

\bibitem{Xue1}
X.~Zhang, Z.~Zheng, W.-Q. Wang, and H.~C. So, ``Joint {DOD} and {DOA} estimation of coherent targets for coprime {MIMO} radar,'' \emph{IEEE Trans. Signal Process.}, vol.~71, pp. 1408--1420, Apr. 2023.

\bibitem{Intro_RF1}
B.~Alexander, S.~Felix, and M.~Igor, ``Practical applications of free-space optical underwater communication,'' in \emph{2021 Fifth Underwater Communications and Networking Conference (UComms)}, Nov. 2021, pp. 1--5.

\bibitem{Intro_RF2}
K.~Takizawa, R.~Suga, T.~Matsuda, and F.~Kojima, ``Underwater {MIMO} communications by {RF} signals: Capacity analysis, simulations, and experiment,'' in \emph{2021 Joint European Conference on Networks and Communications \& 6G Summit (EuCNC/6G Summit)}, Jun. 2021, pp. 460--465.

\bibitem{Intro_RF3}
A.~Hajiaghajani, P.~Rwei, A.~H. Afandizadeh~Zargari, A.~R. Escobar, F.~Kurdahi, M.~Khine, and P.~Tseng, ``Amphibious epidermal area networks for uninterrupted wireless data and power transfer,'' \emph{Nature Communications}, vol.~14, Nov. 2023.

\bibitem{Intro_RF4}
C.~Zeng, J.-B. Wang, C.~Ding, M.~Lin, and J.~Wang, ``{MIMO} unmanned surface vessels enabled maritime wireless network coexisting with satellite network: Beamforming and trajectory design,'' \emph{IEEE Trans. Commun.}, vol.~71, no.~1, pp. 83--100, Jan. 2023.

\bibitem{WCM}
Y.~Li, J.~Xu, N.~Khachatrian, and M.-S. Alouini, ``Implementation of an {IEEE} 802.11ax-based maritime mesh network in the red sea,'' \emph{IEEE Wirel. Commun.}, pp. 1--7, Mar. 2025.

\bibitem{Intro_UOWC1}
A.~Celik, I.~Romdhane, G.~Kaddoum, and A.~M. Eltawil, ``A top-down survey on optical wireless communications for the internet of things,'' \emph{IEEE Commun. Surv. Tutor.}, vol.~25, no.~1, pp. 1--45, Feb. 2023.

\bibitem{Intro_UOWC2}
M.~Salman, J.~Bolboli, R.~P. Naik, and W.-Y. Chung, ``Aqua-sense: Relay-based underwater optical wireless communication for {IoUT} monitoring,'' \emph{IEEE Open J. Commun. Soc.}, vol.~5, pp. 1358--1375, Feb. 2024.

\bibitem{Xue3}
X.~Zhang, Z.~Zheng, W.-Q. Wang, and H.~C. So, ``{DOA} estimation of coherent sources using coprime array via atomic norm minimization,'' \emph{IEEE Signal Process. Lett.}, vol.~29, pp. 1312--1316, May. 2022.

\bibitem{Intro_UOWC3}
J.~Zhang, S.~Wang, Z.~Ma, G.~Gao, Y.~Guo, F.~Zhang, S.~Huang, and J.~Zhang, ``Long-term and real-time high-speed underwater wireless optical communications in deep sea,'' \emph{IEEE Commun. Mag.}, vol.~62, no.~3, pp. 96--101, Mar. 2024.

\bibitem{Intro_UOWC4}
F.~Feng, Y.~Liu, K.~Zhang, H.~Yang, B.-R. Hyun, K.~Xu, H.-S. Kwok, and Z.~Liu, ``High-power algan deep-ultraviolet micro-light-emitting diode displays for maskless photolithography,'' \emph{Nature Photonics}, vol.~19, Jan. 2025.

\bibitem{Intro_UOWC5}
P.~Sanmartín, F.~Almonacid, M.~A. Ceballos, A.~García-Loureiro, and E.~F. Fernández, ``Wide-bandgap {III-V} materials for high efficiency air and underwater optical photovoltaic power transmission,'' \emph{Solar Energy Materials and Solar Cells}, vol. 266, p. 112662, Mar. 2024.

\bibitem{Intro_UWSN1}
J.~Li, J.~Wu, C.~Li, W.~Yang, A.~K. Bashir, J.~Li, and Y.~D. Al-Otaibi, ``Information-centric wireless sensor networking scheme with water-depth-awareness content caching for underwater {IoT},'' \emph{IEEE Internet Things J.}, vol.~9, no.~2, pp. 858--867, Jan. 2022.

\bibitem{Intro_UWSN2}
N.~Kumari and M.~Sajwan, ``Underwater wireless sensor networks ({UWSN}): Challenges and advances in routing protocols,'' in \emph{2024 12th International Conference on Internet of Everything, Microwave, Embedded, Communication and Networks (IEMECON)}, Jan. 2024, pp. 1--6.

\bibitem{SDPXu}
T.~Xu, Y.~Hu, B.~Zhang, and G.~Leus, ``{RSS}-based sensor localization in underwater acoustic sensor networks,'' in \emph{2016 IEEE International Conference on Acoustics, Speech and Signal Processing {(ICASSP)}}, May. 2016, pp. 3906--3910.

\bibitem{SDSOCP}
S.~Chang, Y.~Li, Y.~He, and Y.~Wu, ``{RSS}-based target localization in underwater acoustic sensor networks via convex relaxation,'' \emph{Sensors}, vol.~19, no.~10, May. 2019.

\bibitem{urick1975principles}
R.~J. Urick, ``Principles of underwater sound-2,'' 1975.

\bibitem{FCUP}
Y.~Li, B.~Mukhopadhyay, and M.-S. Alouini, ``{RSS}-based cooperative localization and transmit power(s) estimation using mixed {SDP-SOCP},'' \emph{IEEE Trans. Veh. Technol.}, pp. 1--6, Jul. 2023.

\bibitem{U1}
T.~L.~N.~Nguyen and Y.~Shin, ``An efficient {RSS} localization for underwater wireless sensor networks,'' \emph{Sensors}, vol.~19, no.~14, Jul. 2019.

\bibitem{U3}
A.~Pourkabirian, F.~Kooshki, M.~H. Anisi, and A.~Jindal, ``An accurate {RSS/AoA}-based localization method for internet of underwater things,'' \emph{Ad Hoc Networks}, vol. 145, p. 103177, Jun. 2023.

\bibitem{CM_S1}
J.~Xian, J.~Ma, X.~Mei, H.~Wu, N.~Saeed, D.~Han, M.~D. Marino, and K.-C. Li, ``Robust coarse-to-fine {3D}-target-localization algorithm for underwater-{IoT}-based networks: Design and performance evaluation under uncertain multi-parameters,'' \emph{IEEE Internet Things J.}, pp. 1--1, Apr. 2025.

\bibitem{mag}
N.~Patwari, J.~Ash, S.~Kyperountas, A.~Hero, R.~Moses, and N.~Correal, ``Locating the nodes: cooperative localization in wireless sensor networks,'' \emph{IEEE Signal Process. Mag.}, vol.~22, no.~4, pp. 54--69, Jul. 2005.

\bibitem{Weighted1}
S.~Tomic, M.~Beko, R.~Dinis, and P.~Montezuma, ``A closed-form solution for {RSS/AoA} target localization by spherical coordinates conversion,'' \emph{IEEE Wirel. Commun. Lett.}, vol.~5, no.~6, pp. 680--683, Dec. 2016.

\bibitem{Weighted2}
S.~Tomic, M.~Beko, and R.~Dinis, ``{3-D} target localization in wireless sensor networks using {RSS} and {AoA} measurements,'' \emph{IEEE Trans. Veh. Technol.}, vol.~66, no.~4, pp. 3197--3210, Apr. 2017.

\bibitem{Intro_AI1}
H.~Lee, K.~Kim, T.~Chung, and H.~Ko, ``Deep learning-based ultra short baseline underwater positioning,'' in \emph{2023 International Conference on Artificial Intelligence in Information and Communication (ICAIIC)}, Mar. 2023, pp. 856--859.

\bibitem{Intro_AI2}
I.~B. Saksvik, A.~Alcocer, and V.~Hassani, ``A deep learning approach to dead-reckoning navigation for autonomous underwater vehicles with limited sensor payloads,'' in \emph{OCEANS 2021: San Diego – Porto}, Feb. 2022, pp. 1--9.

\bibitem{R_AI}
W.~M. Salama, M.~H. Aly, and E.~S. Amer, ``Optimized deep learning/kalman filter-based underwater localization in {VLC} systems,'' \emph{Optical and Quantum Electronics}, vol.~55, Jan. 2023.

\bibitem{R4C1_1}
U.~Draz, A.~Ali, M.~Bilal, T.~Ali, M.~A. Iftikhar, A.~Jolfaei, and D.~Y. Suh, ``Energy efficient proactive routing scheme for enabling reliable communication in underwater internet of things,'' \emph{IEEE Trans. Netw. Sci. Eng.}, vol.~8, no.~4, pp. 2934--2945, Sep. 2021.

\bibitem{R4C1_2}
U.~Draz, T.~Ali, N.~Ahmad~Zafar, A.~Saeed~Alwadie, M.~Irfan, S.~Yasin, A.~Ali, and M.~A. Khan~Khattak, ``Energy efficient watchman based flooding algorithm for {IoT}-enabled underwater wireless sensor and actor networks,'' \emph{ETRI Journal}, vol.~43, no.~3, pp. 414--426, Apr. 2021.

\bibitem{R4C1_3}
T.~Ali, S.~Yasin, U.~Draz, and M.~Ayaz, ``Towards formal modeling of subnet based hotspot algorithm in wireless sensor networks,'' \emph{Wireless Personal Communications}, vol. 107, pp. 1573 -- 1606, Aug. 2019.

\bibitem{R4C1_4}
U.~Draz, T.~Ali, S.~Yasin, S.~Bukhari, M.~S. Khan, M.~Hamdi, S.~Rahman, L.~T. Jung, and A.~Ali, ``An optimal scheme for {UWSAN} of hotspots issue based on energy-efficient novel watchman nodes,'' \emph{Wireless Personal Communications}, vol. 121, pp. 69 -- 94, Nov. 2021.

\bibitem{R4C1_5}
T.~Ali, U.~Draz, S.~Yasin, M.~Hijji, M.~Ayaz, I.~Yasin, and T.~Alhmiedat, ``Optimization and localization based framework for priority-aware node ranking and routing in {IoT}-driven acoustic systems,'' \emph{Ad Hoc Networks}, vol. 176, p. 103893, Sep. 2025.

\bibitem{R4C1_6}
U.~Draz, T.~Ali, S.~Yasin, M.~H. Chaudary, I.~Yasin, M.~Ayaz, and E.-H.~M. Aggoune, ``Hybrid underwater localization communication framework for blockchain-enabled {IoT} underwater acoustic sensor network,'' \emph{IEEE Internet Things J.}, vol.~12, no.~11, pp. 16\,858--16\,885, Jan. 2025.

\bibitem{CM_S2}
M.~Tasnim and R.~H. Ratul, ``Optoacoustic signal-based underwater node localization technique: Overcoming {GPS} limitations without {AUV} requirements,'' in \emph{2023 IEEE Symposium on Wireless Technology \& Applications (ISWTA)}, Sep. 2023, pp. 7--11.

\bibitem{CTUP}
Y.~Li, B.~Mukhopadhyay, J.~Xu, and M.-S. Alouini, ``Experimental validation of cooperative {RSS}-based localization with unknown transmit power, path loss exponent, and precise anchor location,'' \emph{IEEE Trans. Wirel. Commun.}, pp. 1--1, Aug. 2024.

\bibitem{Absorption}
M.~C. Domingo, ``Overview of channel models for underwater wireless communication networks,'' \emph{Physical Communication}, vol.~1, no.~3, pp. 163--182, Sep. 2008.

\bibitem{Bisection}
A.~Beck, P.~Stoica, and J.~Li, ``Exact and approximate solutions of source localization problems,'' \emph{IEEE Trans. Signal Process.}, vol.~56, no.~5, pp. 1770--1778, May. 2008.

\bibitem{GTRS2}
J.~J. MORE, ``Generalizations of the trust region problem,'' \emph{Optimization Methods and Software}, vol.~2, no. 3-4, pp. 189--209, Feb 1993.

\bibitem{Frequency}
\BIBentryALTinterwordspacing
E.~GmbH. (2025, May.) Depth-rated long-range devices. [Online]. Available: \url{https://www.evologics.com/acoustic-modems}
\BIBentrySTDinterwordspacing

\bibitem{SDP_Zou}
Y.~Zou and H.~Liu, ``{RSS}-based target localization with unknown model parameters and sensor position errors,'' \emph{IEEE Trans. Veh. Technol.}, vol.~70, no.~7, pp. 6969--6982, Jun. 2021.

\bibitem{SDP_Wang}
Q.~Wang, Z.~Duan, and F.~Li, ``Semidefinite programming for wireless cooperative localization using biased {RSS} measurements,'' \emph{IEEE Commun. Lett.}, vol.~26, no.~6, pp. 1278--1282, Apr. 2022.

\bibitem{Underwater_ChannelEstimate1}
W.~Jiang and R.~Diamant, ``Long-range underwater acoustic channel estimation,'' \emph{IEEE Trans. Wirel. Commun.}, vol.~22, no.~9, pp. 6267--6282, Sep. 2023.

\bibitem{Underwater_ChannelEstimate2}
X.~Qin and R.~Diamant, ``Joint channel estimation and decoding for underwater acoustic communication with a short pilot sequence,'' \emph{IEEE J. Ocean. Eng.}, vol.~48, no.~2, pp. 526--541, Apr. 2023.

\end{thebibliography}
\bibliographystyle{IEEEtran}

\end{document}